\documentclass[a4paper,fleqn]{cas-sc}
\ExplSyntaxOn
\cs_gset:Npn \__first_footerline: 
{ 
	\group_begin: 
	\small \sffamily 
	Preprint ~ submitted ~ Journal
	\group_end: 
}
\ExplSyntaxOff
\usepackage[authoryear,round]{natbib}
\usepackage{subcaption}
\usepackage{cleveref}
\usepackage{rotating}

\def\tsc#1{\csdef{#1}{\textsc{\lowercase{#1}}\xspace}}
\tsc{WGM}
\tsc{QE}

\begin{document}
\let\WriteBookmarks\relax
\def\floatpagepagefraction{1}
\def\textpagefraction{.001}

\shorttitle{FL-NSE to solve bi-Objective service-oriented TSP}

\shortauthors{S.Abdoune and M.Boulif}

\title [mode = title]{Node-Shift-Encoding Genetic Algorithm with fuzzy-enhanced reference tour to solve the bi-objective service-oriented TSP}  

\author{Souad Abdoune}
\ead{s.abdoune@univ-boumerdes.dz}
\credit{Conceptualizing, Programming, Writing, Analysing}

\author{Menouar Boulif}
\cormark[1]
\ead{boumen7@gmail.com}
\credit{Conceptualizing, Writing, Correcting, Analysing, Supervising, Validating}

\affiliation{organization={LIMOSE Laboratory, University of M'hamed Bougara},
            postcode={35000}, 
            state={Boumerdes},
            country={Algeria}}

\cortext[1]{Corresponding author}

\begin{abstract}
The Travelling Salesman Problem (TSP) remains a key area of research in combinatorial optimization, with applications in logistics, manufacturing, and service delivery. This paper addresses a bi-objective service-oriented TSP in which the clients' ranks in the delivery path matter. Unlike conventional depot-based TSP formulations, the considered problem does not assume a distinguished depot or a fixed tour origin. To address this setting, we adapt the Miller--Tucker--Zemlin (MTZ)-based formulation and derive an original linearization of the resulting model, enabling its solution with off-the-shelf integer linear programming solvers. This adaptation avoids the rigid tour origin imposed by the conventional MTZ formulation, for which fixing the starting node does not affect the tour cost but can affect the objective in a customer-rank-sensitive TSP. To solve this problem, we present a Node-Shift-Encoding (NSE)-based Genetic Algorithm augmented with fuzzy reasoning to update the reference tour throughout the evolutionary process. Experimental evaluation on TSPLIB benchmarks demonstrates that the proposed method achieves improved performance compared with the classical NSE approach.
\end{abstract}


\maketitle

\section{Introduction}
    
The Travelling Salesman Problem (TSP) is a fundamental NP-hard problem in combinatorial optimization, with diverse applications in logistics, transportation, and service systems \cite{Goyal2010}. Classical formulations focus exclusively on identifying a minimum-cost tour that visits a set of cities exactly once and returns to the origin. By concentrating solely on distance or travel path length, these traditional models implicitly assume homogeneous service requirements across all locations.
    
However, this assumption is often unrealistic in modern service-oriented and automated systems, where locations exhibit heterogeneous priorities and service expectations. For instance, in domestic healthcare service delivery, a mobile nurse must visit all patients in a neighbourhood, but specific individuals require earlier attention due to medical urgency, frailty, medication timing, or personal scheduling constraints. Similarly, in robotic inspection coverage, a mobile unit must inspect all designated viewpoints once, yet certain sites demand priority due to operational risk, critical status, or sensor requirements.
    
When operational settings demand that specific locations be serviced at particular positions along a route, failing to account for these priorities produces tours that are optimal in terms of distance, yet inadequate from a service perspective. Unlike prize-collecting or subset-selection variants where low-priority nodes can be omitted, many real-world applications require full coverage while making the service order critical.
    
This challenge motivates a service-oriented, bi-objective extension of the TSP that balances travel efficiency against position-based service priorities. In this framework, all nodes remain mandatory, and each location is associated with a target or maximal visit rank. Any deviation from the assigned maximal visit rank induces a dissatisfaction cost. By simultaneously minimizing overall travel distance and total customer dissatisfaction, the model bridges the gap between operational cost reduction and high-quality service delivery.

Genetic Algorithms (GAs) have proven highly effective for the TSP and its variants since the pioneering work of Holland~\cite{Holland1975} and Goldberg~\cite{Goldberg1989}. Their success depends critically on the encoding scheme, which must preserve solution feasibility under standard genetic operators. Early approaches relied on \emph{Path Representation}, where a tour is stored as an ordered list of cities. While intuitive, this scheme requires specialized operators such as Partially Mapped Crossover (PMX), Order Crossover (OX), or Cycle Crossover (CX) to maintain permutation validity, together with tailored mutation operators (e.g., swap, insertion, or inversion) that supersede classical one-point bit-flip mutation; without these, dedicated repair mechanisms become necessary~\cite{larranaga1999genetic}. Table \ref{tab:comparative} draws a comparison between the proposed work and some existing evolutionary approaches.
    
\begin{table}[htbp]
    \centering
    \caption{Comparison to existing evolutionary approaches for the TSP.}
    \label{tab:comparative}
    \resizebox{\textwidth}{!}{%
    \begin{tabular}{l c l c c c c p{10cm}}
    \hline
    \textbf{Ref.} &
    \textbf{Year} &
    \textbf{Encoding} &
    \makecell{\textbf{Compatible with}\\
    \textbf{std. operators}} &
    \makecell{\textbf{Multi-}\\\textbf{objective}} &
    \makecell{\textbf{Math. Prog.}\\\textbf{approach}}&
    \makecell{\textbf{Fuzzy logic}} &
    \textbf{Notes} \\ 
    
    \midrule
    
    \cite{Goldberg1985} &
    1985 &
    Path representation &
    $\boldsymbol{\times}$ &
    $\boldsymbol{\times}$ &
    $\boldsymbol{\times}$ &
    $\boldsymbol{\times}$ &
    Classical permutation encoding. introduce Partially Mapped Crossover (PMX)  \\        
    
    \cite{Grefenstette1985} &
    1985 &
    Adjacency representation &
    $\boldsymbol{\times}$ &
    $\boldsymbol{\times}$&
    $\boldsymbol{\times}$ &
    $\boldsymbol{\times}$ &
    Represents each node by a list of cities. Uses specialised heuristic crossover \\  
   
    \cite{Homaifar1992} &
    1992 &
    Adjacency matrix &
    \checkmark &
    $\boldsymbol{\times}$ &
    $\boldsymbol{\times}$ &
    $\boldsymbol{\times}$ &
    Encodes tours via $n \times n$ adjacency matrix. Conventional crossover is applicable.  \\
    
    \cite{Riazi2019} &
    2019 &
    Double-chromosome &
    \checkmark &
    $\boldsymbol{\times}$ &
    $\boldsymbol{\times}$ &
    $\boldsymbol{\times}$ &
    Proposes a double-chromosome encoding for the TSP. Two coordinated chromosomes represent the candidate tour in order to preserve permutation feasibility during evolution. \\  
    
    \cite{fernandez2021} &
    2021 &
    Ordinal representation &
    \checkmark &
    $\boldsymbol{\times}$ &
    $\boldsymbol{\times}$ &
    $\boldsymbol{\times}$ &
    Investigates a greedy single-point crossover operator for the ordinal representation.  \\   
      
    \cite{Boulif2024} &
    2024 &
    Node-Shift Encoding &
    \checkmark &
    $\boldsymbol{\times}$ &
    \checkmark &
    $\boldsymbol{\times}$ &
    Proposes Node Shift Encoding based on a static reference tour. Standard genetic operators can be used without repair. \\
     
    \textbf{This study} &
    --- &
    Node-Shift Encoding &
    \checkmark &
    \checkmark &
    \checkmark &
    \checkmark &
    Extends NSE to a bi-objective service-oriented TSP. The objectives represent tour cost and customer-priority dissatisfaction, while fuzzy reasoning is used to dynamically update the reference tour. \\ 
    \bottomrule
    
\end{tabular}%
}
\end{table}
    
One of the most promising encoding mechanisms is the Node Shift Encoding (NSE) \cite{Boulif2024}. NSE is known to ensure solution feasibility, as it allows standard bio-inspired crossover and mutation operators to be applied without requiring solution repair mechanisms. One of the limitations of classical NSE, however, is that this scheme use a single, static reference tour as a basis, which can limit exploration ability of the algorithm and cause premature convergence.
        
To overcome this limitation, we propose an adaptive evolutionary framework that utilizes a Fuzzy Logic (FL) based mechanism to dynamically update the reference tour. Our approach detects the stagnation of evolution process and updates the reference tour thanks to a fuzzy inference system. Using an adaptive reference tour allows to explore a larger region of feasible solutions. 
        
The effectiveness of the proposed technique is tested on benchmark instances from TSPLIB library. Experimental results show consistent improvements over classical NSE in terms of solution quality and robustness. 
           
The remainder of this paper is organized as follows: Section ~\ref{sec:tsp} reviews the classical TSP. The next section presents the bi-objective formulation. Section~\ref{sec:NSE} describes the Node-Shift Encoding scheme and the genetic algorithm framework. In section ~\ref{sec:method} we describe the proposed FL-NSE approach. Section ~\ref{sec:experiments} details the experimental setup followed by the computational results in Section ~\ref{sec:results} and a discussion of the adaptive mechanism in Section ~\ref{sec:discussion}. Finally, Section ~\ref{sec:conclusion}  concludes the paper and suggests future research directions.

\section{Service-oriented Travelling Salesman Problem}
\label{sec:tsp}
	
The TSP is a canonical NP-hard problem in combinatorial optimization, as its decision variant reduces from the Hamiltonian Cycle problem, requiring the identification of a minimum-cost tour that traverses each node of a directed graph exactly once. While significant progress has been achieved regarding its resolution, solving large-scale problems remains highly complex.
	
Despite the classic formulation of the problem being purely oriented towards distance minimization, many practical cases require incorporating additional operational constraints along with customer demands. As a result, numerous TSP variants were introduced with various extensions involving service criteria. One of the well-researched TSP variants involves incorporating customer satisfaction considerations.
	
For example, some works introduce time windows during which every node should be visited, referred to as the Traveling Salesman Problem with Time Windows (TSPTW). It results in additional temporal constraints that substantially increase the difficulty of TSP solutions \cite{Jeffrey2007, Papalitsas2019}.
	
The Prize-Collecting TSP (PCTSP) is another important extension, in which each node is associated with a reward and a penalty. In this case, a problem solution seeks to balance the travel costs with collecting maximum rewards \cite{Goemans1995, Bienstock1993}.
	
More recently, variants such as the Traveling Salesman Problem with Priority Prizes (TSPPP) explicitly incorporate customer preferences by associating rewards with the visiting order, reflecting the importance of service prioritization \cite{Pureza2018}.
	
Thus, current extensions demonstrate that pure minimization of travel distances is not sufficient when modeling service-oriented applications of TSP. Instead, modern applications require balancing operational efficiency with customer satisfaction and service quality.
	
In this paper, a new \textit{service-oriented} TSP is proposed, in which soft position constraints are explicitly modelled. Every client is associated to a maximum visit rank, and deviations from this rank induce a dissatisfaction cost. The resulting formulation is naturally expressed as a bi-objective optimization problem, aiming to simultaneously minimize total travel distance and overall customer dissatisfaction.
	
The formal mathematical model of the proposed problem is detailed in the following section.
	
\section{Problem formulation}
	\label{sec:problem}
	
In the graph-theoretic model of the Traveling Salesperson Problem (TSP), a routing instance is mapped onto a weighted {undirected} graph $G = (V, E)$, where each node $i \in V$ represents a customer and each arc $(i, j) \in E$ denotes a travel leg. A weight function $c: E \to \mathbb{R}^+$ assigns a non-negative travel cost $c_{ij}$ to each edge. Solving the classic TSP thus reduces to finding a minimum-cost Hamiltonian cycle:
\begin{equation}
	f(G) = \sum_{(i, j) \in E} c_{ij}
\end{equation}	
that visits every node exactly once before returning to the origin.
	
In this paper, we extend this classic formulation to a bi-criteria variant that jointly accounts for routing efficiency and  position-based service requirements. Using a complete {undirected} graph $G = (V, E)$ with $V = \{1, \dots, n\}$, each node $i \in V$ is additionally associated with a preferred maximum visit position $p_i \in \{1, \dots, n\}$. The problem is formulated as an integer linear program using the Miller--Tucker--Zemlin (MTZ) \cite{Miller1960} subtour elimination constraints.
	
\subsection{Preliminary MTZ-based modelling}
We first define two decision variables:
\begin{itemize}
\item $x_{ij} \in \{0,1\}$: equals 1 if edge $(i,j)$ is selected in the tour, and 0 otherwise
\item $u_i \in \{2, \dots, n\}$: position of node $i$ in the tour, where node 1 always designates the starting point of the tour ($u_1=1$).
\end{itemize}
		
Using these variables, we model position requirements as soft constraints. The dissatisfaction associated with node $i$ is defined as:
\begin{equation}
	\label{delta}
	\delta_i = \max(0, u_i - p_i), \quad \forall i \in V\setminus \{1\}
\end{equation}
	
This formulation penalizes late visits while allowing flexibility in the optimization process.
	
Concerning the critaria, the problem naturally involves two conflicting objectives:
\begin{equation}
	\min D = \sum_{i \in V} \sum_{\substack{j \in V \\ j \neq i}} c_{ij} x_{ij} 
\end{equation}
\begin{equation}
	\label{2nd_objective}
	\min C = \sum_{i \in V \setminus \{1\}} \delta_i
\end{equation}
	
\textbf{Remark:} While our approach seemingly aligns with the priority-based TSP introduced by Schmitz and Niemann~\cite{Schmitz2007}, a closer examination reveals critical distinctions between the two models. First, Schmitz and Niemann derive priorities implicitly from instance labels: cities are assumed to be labeled according to a priority rule (the city with highest/lowest priority has label~$1$/$n$), so that city~$i$ is expected at position~$i$ in the tour. Priorities are thus unique and totally ordered by construction, and no additional priority parameters are required. Our model, by contrast, assigns an explicit maximum rank~$p_i$ to each client, which means two different customers may share the same preferred position and the optimizer must actively resolve these conflicts. Second, the formulation in~\cite{Schmitz2007} is based on a permutation representation of tours without an explicit mathematical programming model, whereas our approach is grounded in a modified Miller--Tucker--Zemlin (MTZ) formulation that can be solved directly using off-the-shelf ILP solvers. This modification, detailed below, removes the conventional dependence on a fixed city ordering (i.e., the implicit assumption that city~$1$ is visited first) and enables the model to handle arbitrary city sequences without positional bias.\\
	
Besides, in this work the two considered objectives are combined using a weighted sum aggregation:
\begin{equation}
	\min F = w_1 D + w_2 C
\end{equation}
where $w_1, w_2 \geq 0$ $(w_1 + w_2 = 1)$ are weighting parameters reflecting the relative importance of each criterion.
	
To be accepted, each solution must adhere to the following constraints:
	
\begin{enumerate}
\item Each customer is visited exactly once:
\begin{equation}
	\label{custVisitIn}
	\sum_{\substack{j \in V \\ j \neq i}} x_{ij} = 1, \quad \forall i \in V
\end{equation}
\begin{equation}
	\label{custVisitOut}
	\sum_{\substack{i \in V \\ i \neq j}} x_{ij} = 1, \quad \forall j \in V
\end{equation}
\item Subtour elimination constraints (MTZ):
\begin{equation}
	\label{MTZ1}
	u_i - u_j + n x_{ij} \le n - 1, \quad \forall i,j \in V\setminus \{1\}, i \neq j
\end{equation}
\begin{equation}
	\label{MTZ2}
	2 \le u_i \le n, \quad \forall i \in V \setminus \{1\}
\end{equation}
\item Binary constraint:
\begin{equation}
	\label{xBinary}
	x_{ij} \in \{0,1\}, \quad \forall i,j \in V, i \neq j
\end{equation}
\end{enumerate}
		
\subsection{Origin-independent formulation} \label{origin_depend_form}
Standard MTZ-based formulations anchor the sequence origin by fixing $u_1 = 1$. In rank-sensitive or preference-driven objectives, this restriction implicitly excludes all solutions that do not initiate at node 1, leaving a significant portion of the valid solution space inaccessible to the optimization model. To resolve this issue, we construct an augmented graph $G'=(V',E')$ by adding a dummy node (indexed as node 1) linked to all original vertices $V$ via zero-weight edges, with all original vertex indices shifted by one.
	
Without further modification, searching for an optimal tour on $G'$ reduces to identifying an optimal Hamiltonian path over the original node set in $G$, rather than a complete Hamiltonian cycle. To restore tour closure, we introduce an auxiliary binary variable $z_{ij} \in \{0,1\}$ that identifies the implicit connection between the termination node $i$ and the initiation node $j$ of the open path. The variable's behavior is governed by the following structural constraint:
\begin{equation}
	\label{missingEdgeCtrt}
	z_{ij} \geq x_{1i}+x_{j1}-1, \quad \forall i,j \in V' \setminus \{1\}
\end{equation}

The travel distance objective (Eq.~\ref{2nd_objective}) is updated to explicitly include the cost of this closure edge:
\begin{equation}
	\label{new1st_objective}
	\min D' = \sum_{i \in V'\setminus \{1\}} \sum_{\substack{j \in V' \setminus \{1\} \\ j \neq i}} c_{ij} x_{ij} 
	+ \sum_{i \in V'\setminus \{1\}} \sum_{\substack{j \in V' \setminus \{1\} \\ j \neq i}} c_{ij} z_{ij} 
\end{equation}
	
By combining Constraint~\eqref{missingEdgeCtrt} with the optimization direction of Eq.~\eqref{new1st_objective}, $z_{ij}$ is forced to 1 if and only if node $i$ is the final customer visited and node $j$ is the first customer visited in the Hamiltonian path, thereby tying the endpoint to the startpoint and forming the complete Hamiltonian cycle relative to $G$.\\
	
Furthermore, the dissatisfaction defined by Eq. \eqref{delta} is updated to compensate the artificial positional bias among customer ranks induced by the dummy node:
\begin{equation}
	\label{newDelta}
	\delta_i = \max(0, u_i - p_i-1), \quad \forall i \in V' \setminus \{1\}
\end{equation}
	
\begin{figure}[htbp]
	\centering
	\includegraphics[width=0.55\textwidth]{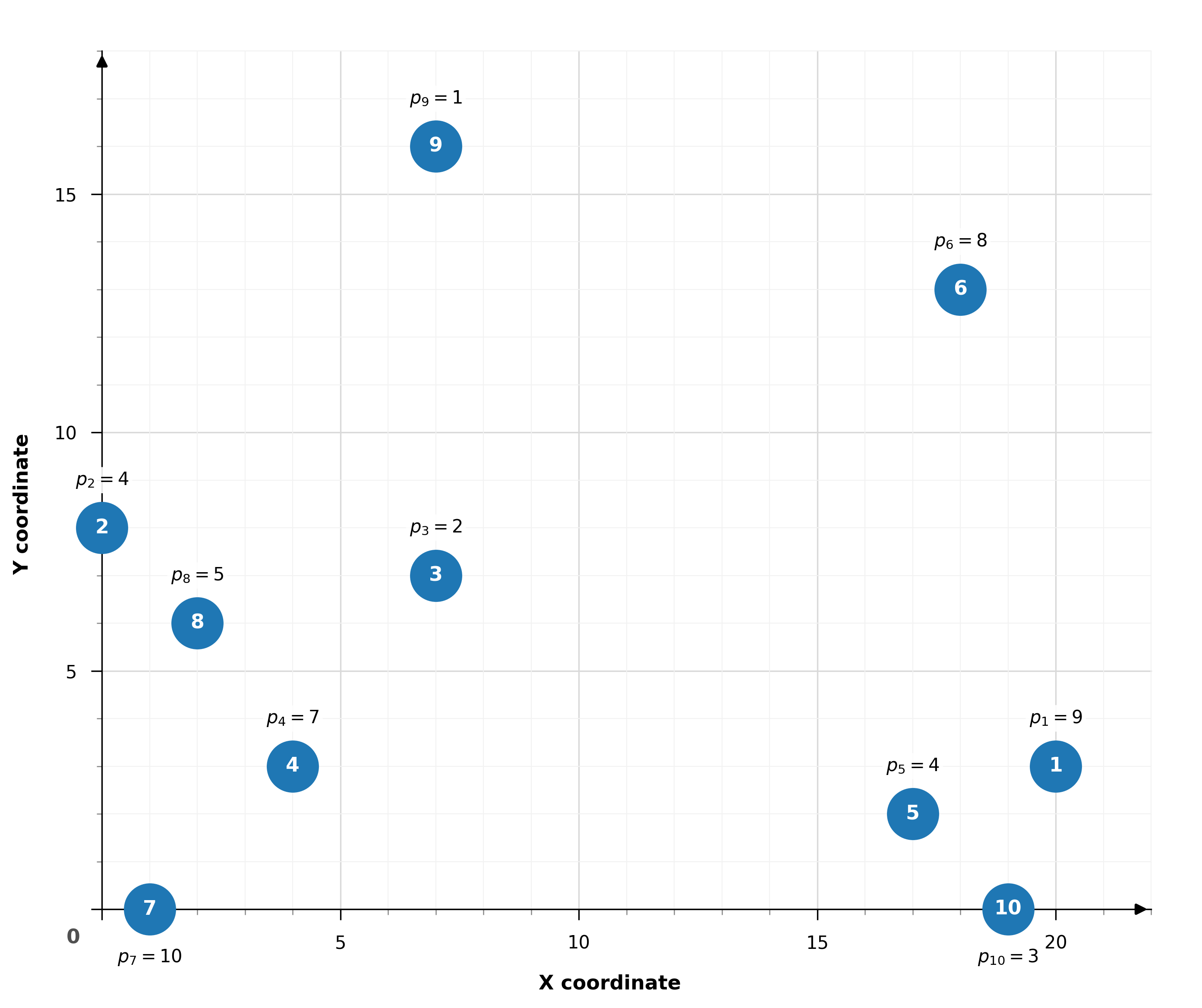}
	\caption{Illustrative instance with $n=10$ customers.}
	\label{fig:instance}
\end{figure}

To provide a clear concrete illustration of this origin-independent transformation, let us examine a simple example.
Consider a bi-objective instance with $n=10$ customers distributed in the Euclidean plane (Figure~\ref{fig:instance}). Travel costs $c_{ij}$ equal the Euclidean distances; preferred maximum ranks $p_i$ are indicated next to each node. We set $w_1=w_2=0.5$, $d_{\max}=160$, $c_{\max}=47$.
	
\begin{figure}[htbp]
	\centering
	\includegraphics[width=0.55\textwidth]{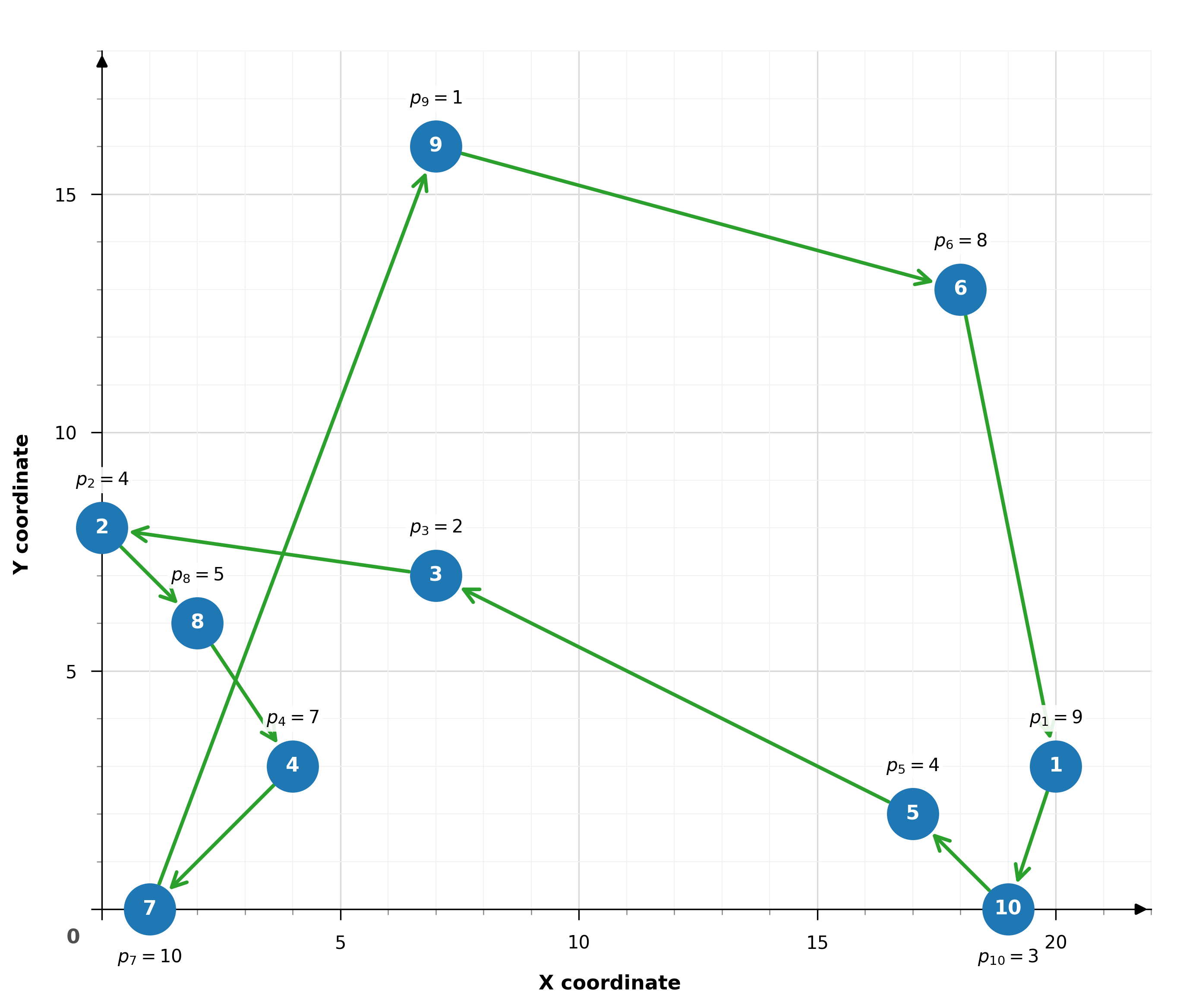}
	\caption{Optimal tour under the standard MTZ formulation with fixed origin at node~$1$.}
	\label{fig:mtz}
\end{figure}

Under the standard MTZ formulation, the origin is anchored at node~$1$ ($u_1=1$). The solver returns the Hamiltonian cycle $	1 \to 10 \to 5 \to 3 \to 2 \to 8 \to 4 \to 7 \to 9 \to 6 \to 1 $, depicted in Figure~\ref{fig:mtz}. This tour yields distance $D=73.61$ and dissatisfaction $C=14$. Because customer~$1$ (with a loose {rank constraint} $p_1=9$) is forced to occupy the first position, the highly constrained customer~$9$ ($p_9=1$) is relegated to position $u_9=9$, incurring a penalty $\delta_9=8$. The fitness is $F=0.37897$.
	
\begin{figure}[htbp]
    \centering
	    \begin{subfigure}[b]{0.49\textwidth}
        \centering
	        \includegraphics[width=\textwidth]{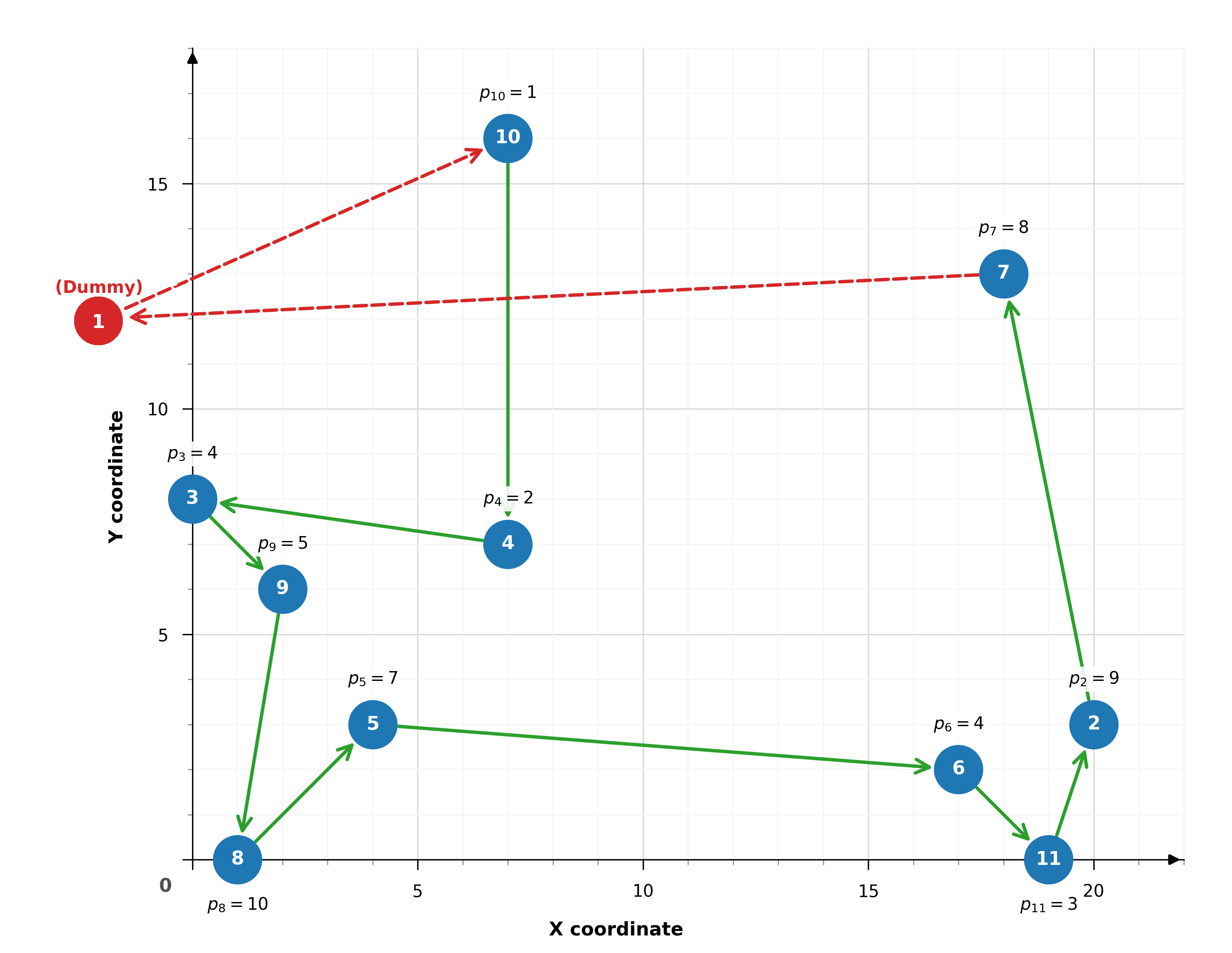}
	        \caption{Augmented graph with dummy node.}
	        \label{fig:dummy}
	    \end{subfigure}
	    \hfill
	    \begin{subfigure}[b]{0.47\textwidth}
	        \centering
	        \includegraphics[width=\textwidth]{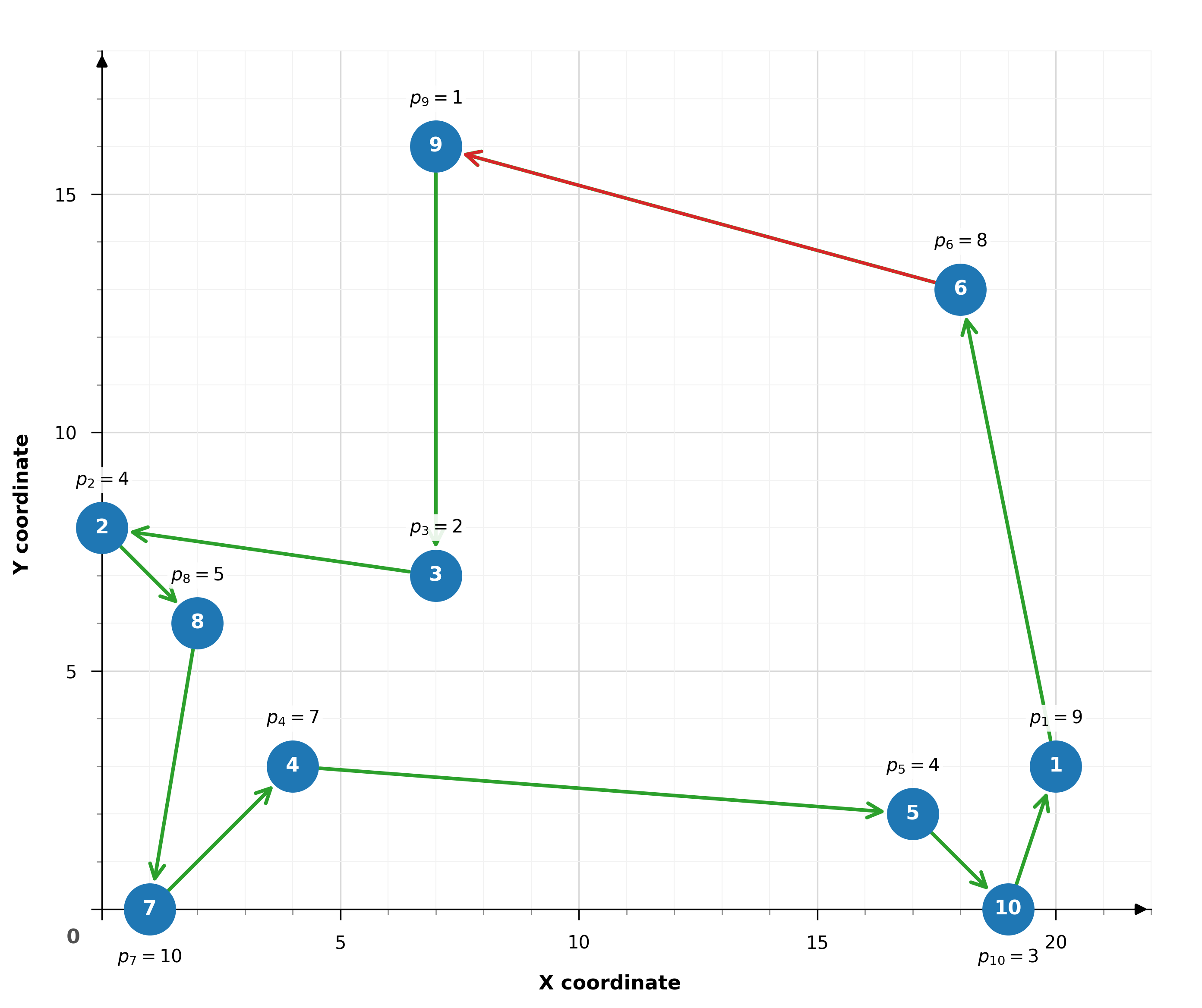}
	        \caption{Restored tour over $V$.}
	        \label{fig:nodummy}
	    \end{subfigure}
    \caption{Optimal tour under the origin-independent formulation}
    \label{fig:dummy-combined}
\end{figure}

Under the dummy-node formulation on $V'$, the optimal Hamiltonian path is	$1 \to 10 \to 4 \to 3 \to 9 \to 8 \to 5 \to 6 \to 11 \to 2 \to 7 \to 1 $ where node~$1$ is the dummy vertex (Figure~\ref{fig:dummy}). Mapping this path back to the original vertex set $V$ and restoring the implicit closing edge $z_{10,7}=1$ yields the equivalent cyclic tour (Figure~\ref{fig:nodummy}): $9 \to 3 \to 2 \to 8 \to 7 \to 4 \to 5 \to 10 \to 1 \to 6 \to 9$ .
The total distance becomes $D'=69.85$ and the compensated dissatisfaction drops to $C'=10$, giving $F=0.32466$.
Notably, this sequence achieves a superior objective value because the standard MTZ formulation restricts the search space exclusively to tours initiating at node~$1$(Table~\ref{tab:comparison}).
\begin{table}[h!]
\centering
\caption{Basic MTZ \emph{vs.} origin-independent formulation.}
\label{tab:comparison}
\begin{tabular}{@{}lcc@{}}
\toprule
 & Basic MTZ & Origin-independent \\ 
\midrule
Tour & $(1,10,5,3,2,8,4,7,9,6,1)$ & $(9,3,2,8,7,4,5,10,1,6,9)$ \\
Distance & $73.61$ & $69.85$ \\
Dissatisfaction & $14$ & $10$ \\
Fitness $F$ & $0.37897$ & $0.32466$ \\
\bottomrule
\end{tabular}
\end{table}

\subsection{Linearization}
\label{linFormul}
Equation~\ref{newDelta} can be linearized as follows. We introduce an integer decision variable:
\begin{equation}
	y_{i} \in \mathbb{N}, \quad \forall i \in V' \setminus \{1\}
\end{equation}
and the following constraint:
\begin{equation}
	\label{linObj2}
	y_i \geq u_i - p_i-1, \quad \forall i \in V' \setminus \{1\}
\end{equation}
We then replace the second objective (Eq.~\ref{2nd_objective}) with:
\begin{equation}
	\label{new2nd_objective}
	\min C' = \sum_{i \in V'\setminus \{1\}} y_i
\end{equation}
This revised objective forces $y_i = \max(0, u_i - p_i-1)$ for every optimal solution, effectively driving $y_i$ to $u_i - p_i-1$ when positive and to zero otherwise.\\
	
Ultimately, combining the bi-objective functions \cref{new1st_objective,new2nd_objective} with constraints \cref{custVisitIn,custVisitOut,MTZ1,MTZ2,missingEdgeCtrt,linObj2} yields an integer linear programming formulation. Here, $V$ is replaced by the augmented vertex set $V'$ where appropriate, alongside standard variable domain definitions. The resulting model can be directly solved using off-the-shelf optimization solvers.
	
\section{ NSE-based Genetic Algorithm}
\label{sec:NSE}	
Genetic Algorithms (GAs) \cite{Holland1975, Goldberg1989} are population-based metaheuristic optimization techniques based on the natural evolution mechanism. GAs have been widely applied for solving the TSP, due to their efficiency in exploring the solution space. The efficiency of GAs depends critically on the solution representation approach and the genetic operators, especially in terms of maintaining feasibility and preserving the valuable structure of the solutions.
		
In this work, a tailored GA is developed based on the Node Shift Encoding (NSE) \cite{Boulif2024}, allowing efficient manipulation of solutions while preserving feasibility. Each individual in the population encodes a feasible tour as a shift vector relative to a reference tour. The evaluation of individuals is performed after decoding the NSE representation into a complete tour \cite{Abdoune2026}.
	
\subsection{Node Shift Encoding}
	
NSE represents a solution as a sequence of shift operations applied to a reference tour, rather than directly encoding permutations. This representation ensures that all generated solutions remain feasible without requiring repair mechanisms, which is a common limitation in permutation-based encodings.
	
Formally, let $R = (r_1, r_2, \dots, r_n)$ denote a reference tour, which is a permutation of the $n$ vertices. A solution is encoded by a shift vector $S = (s_2, s_3, \dots, s_n)$ of length $n-1$, where each element $s_i \in [0, n-2]$ specifies how node $i$ is repositioned relative to its position in $R$, while the first node remains fixed. The decoding process reconstructs a feasible tour by sequentially applying the shift operations to the reference tour \cite{Abdoune2026}.

\subsection{Genetic operators and evolutionary scheme}
	
To explore the NSE space, the proposed GA integrates simple steps :
\begin{itemize}
	\item \textbf{Population initialization:} The initial population of shift vectors is generated uniformly at random within the valid bounds $s_i \in [0, n-2]$.
	\item \textbf{Selection:} A \textit{Roulette Wheel Selection} scheme is adopted, paired with an elitist replacement strategy to preserve the overall best solution across generations.
	\item \textbf{Crossover operator:} A single-point vector crossover is applied directly to the shift vectors $S$. Because every position $s_i$ remains within its valid integer interval, offspring vectors are guaranteed to decode into valid tours.
	\item \textbf{Mutation operator:} A localized shift-perturbation mutation randomly modifies components $s_i$ to introduce local path variations without disrupting global sequence structures.
		
\end{itemize}
		
\section{Approach}
\label{sec:method}
We propose an adaptive hybrid optimization framework that uses an NSE-based Genetic Algorithm and a Fuzzy Logic control mechanism. This approach aims to address the limitations of the static encoding strategy of NSE. Compared to classical permutation representations, NSE offers a significant advantage: it preserves feasibility under genetic operators. However, its performance heavily depends on the choice of the reference tour. Therefore, through controlled perturbations of the reference solution, we aim to achieve a more balanced exploration/exploitation of the solution space. This motivation underlies the proposed work.
     
\subsection{General framework}

Unlike the classical NSE-based method that relies on a fixed reference tour throughout the evolutionary process, the proposed approach introduces a dynamic adaptation mechanism that updates the reference tour during the search. This adaptation is guided by a fuzzy inference system that evaluates candidate solutions based on multiple criteria. The objective is to enhance the exploration capability of the algorithm while maintaining solution feasibility and improving convergence behaviour.

The proposed framework is composed of three main components: Node Shift Encoding for solution representation, a tailored Genetic Algorithm for search exploration, and a Fuzzy Logic controller for adaptive reference tour update.
		
The GA starts with an initial population of random NSE chromosomes. The fitness of each solution is calculated by first decoding the chromosome using a randomly generated initial reference tour. The fitness function of the GA is defined as the normalized weighted sum of the two objective functions defined in Section \ref{sec:problem}. Let $D(s)$ represent the total travel distance and $C(s)$ represent the total customer dissatisfaction of the solution $s$. The fitness function is defined as follows:
	
\begin{equation}
Fitness(s) = w_1 \cdot \frac{D(s)}{d_{max}} + w_2 \cdot \frac{C(s)}{c_{max}}
\end{equation}
	
Where:
\begin{itemize}
	\item $w_1$ and $w_2$ represent the weighting coefficients for the importance of each objective function. 
	\item $d_{max}$ is a travel distance  upper bound. It is computed using farthest neighbor heuristic. It starts from a node and chooses the node that is farthest from the set of unvisited nodes. It continues until all nodes are visited and ends at the starting node.
	\item $c_{max}$ is a dissatisfaction degree upper bound. It is computed by considering the worst case scenario. For a customer $i$ whose preferred location is $p_i$, the greatest dissatisfaction arises when customer $i$ is assigned to the final location, denoted as $n$. This is mathematically represented as: 
	
	\begin{equation}
	 c_{max} = \sum_{i \in V} \max(0, n - p_i) 
	\end{equation}
\end{itemize}
		
In the selection step, a proportion of individuals is selected using the roulette wheel mechanism \cite{Goldberg1989}, favouring individuals with better fitness, while the remaining individuals are generated randomly to enhance diversity.
	
We apply One-point crossover operator to the selected parents (shift vectors) chosen randomly without replacement. Offspring directly replace their parents. Then, a one-point mutation is applied to random individuals by changing a random position in the shift vector. This process is repeated until reaching a maximum number of generations.
	
To preserve high-quality solutions, elite chromosomes are maintained and reinjected into the population at the end of each generation using a tournament. That is, they will replace randomly chosen individuals only if they provide better fitness.
	
\subsection{Fuzzy Logic}
	
Fuzzy Logic (FL), first introduced by Zadeh in 1965 \cite{Zadeh1965} and later extended by the introduction of linguistic variables and approximate reasoning \cite{Zadeh1973,Zadeh1975}, is a highly flexible mathematical tool for dealing with and manipulating uncertainties and imprecise data. Unlike traditional logic, FL allows membership of elements to more than one set to a certain degree by defining membership functions over the interval $[0,1]$. Such a feature of FL is highly beneficial for dealing with complex decision-making processes of optimization problems where exact modeling is difficult \cite{Ross2010, Herrera2003}.
	
Regarding the application of FL with meta-heuristic optimization techniques, FL has been effectively employed as a control mechanism to assist the search process of the optimization algorithm. Such a feature of FL is highly beneficial and has been effectively employed to assist the search process of the Genetic Algorithm (GA) \cite{Boulif2008,Syzonov2024}.
	
FL's functionality is based on the following steps: fuzzification, inference, and defuzzification. In the fuzzification step, crisp input values are mapped to fuzzy sets through membership function definitions of various linguistic terms such as low, medium, and high. In the inference step, a set of fuzzy if-then rules is used to mimic the decision process. The defuzzification step maps the fuzzy output to a crisp value through the application of multiple methods such as the centroid (center of gravity) method \cite{Mamdani1975,Ross2010}.
	
\subsubsection{Fuzzy system design}
\label{Fuzzy_syst}
In this work, FL helps the decision mechanism to dynamically update the reference tour used in the NSE based GA.
	
The idea is to monitor the search progress and to activate the fuzzy system when  there is stagnation. Rather than continuing the search around a potentially limiting reference tour, the algorithm will make an intelligent choice of a reference tour from the current population. The fuzzy system is activated when no significant improvement is detected in the best fitness value over a certain number of consecutive generations.

\begin{enumerate}
	\item \textbf{Linguistic variables}\\
	In order to use fuzzy reasoning we define the following linguistic variables. The two first are input variables and the later is the output.
	\begin{itemize}
		\item Route Distance (\textit{RD}): Reflects the quality of the tour in terms of travel distance.
		\item Customer Dissatisfaction (\textit{CD}): Measures how well the tour satisfies { the assigned position requirements}.
		\item Selection score (\textit{SS}): If the score is high, the solution is more likely to be selected as the new reference tour.
	\end{itemize}
	
	The linguistic variable values are defined using simple and interpretable terms (see table \ref{tab:linguistic_var}). The associated membership functions are shown in Fig.\ref{fig:membership_function}.
			
	\begin{table}[htbp]
	\centering
	\caption{Linguistic variables and their membership function parameters}
	\label{tab:linguistic_var}
	\begin{tabular}{@{}ll@{}}
	\hline
	\textbf{Variable} & \textbf{Values}\\
	\hline
	Route Distance (\textit{RD}) & Small , Long \\
	Customer Dissatisfaction (\textit{CD}) & Low , High \\
	Selection Score (\textit{SS}) & Low , Medium , High \\
	\hline
	\end{tabular}
	\end{table}
	\begin{figure}[htbp]
	\centering
	\includegraphics[width=0.95\linewidth]{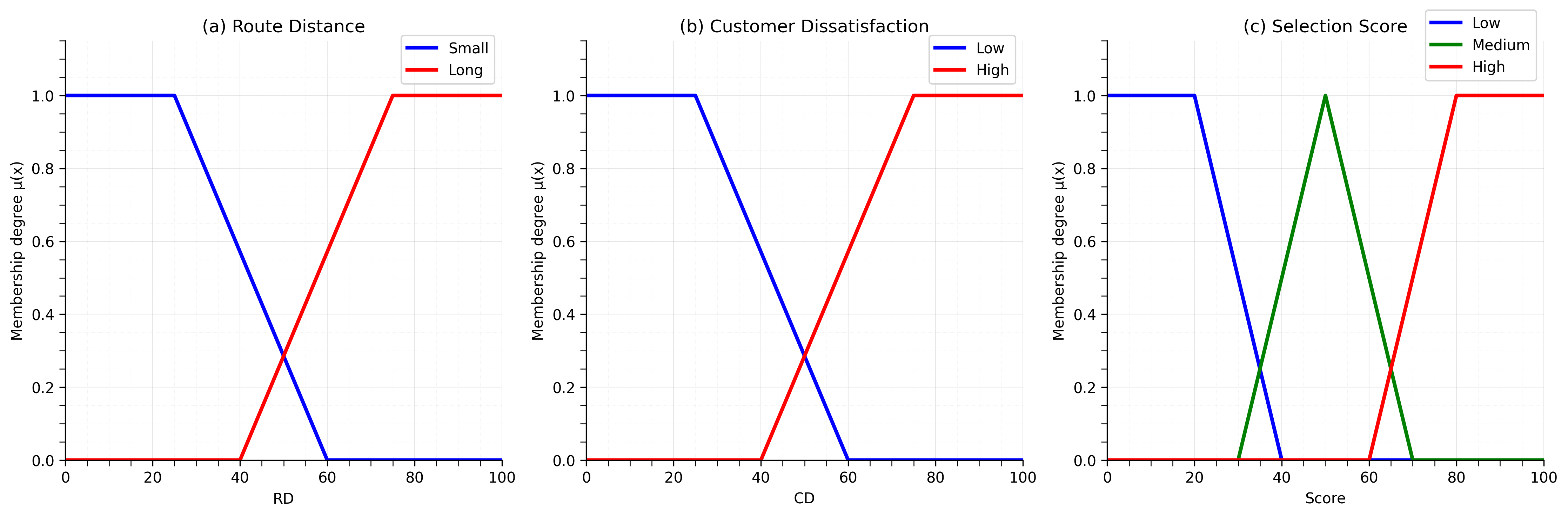}
	\caption{Fuzzy membership functions.}
	\label{fig:membership_function}
	\end{figure}
	\item \textbf{Rule base} \\
	
	The fuzzy decision-making process is guided by four rules, structured as follows:
	
	\begin{itemize}
	    \item If \textit{RD} is Small and \textit{CD} is Low, then \textit{SS} is High.
	    \item If \textit{RD} is Long and \textit{CD} is Low, then \textit{SS} is Medium.
	    \item If \textit{RD} is Small and \textit{CD} is High, then \textit{SS} is Medium.
	    \item If \textit{RD} is Long and \textit{CD} is High, then \textit{SS} is Low.
	\end{itemize}
	
	The rules emphasize solutions that achieve a good balance between the two objectives rather than focusing on maximizing one objective at the expense of the other.
	
	\item \textbf{Inference and defuzzification} \\
	We adopt Mamdani's fuzzy inference mechanism \cite{Mamdani1975} to build the control system. The firing strength of each rule is evaluated using the minimum ($\min$) operator, while rule aggregation is performed via the maximum ($\max$) operator. Finally, the crisp output score is computed using the Center of Gravity defuzzification method.
	
	\item \textbf{Reference tour update} \\
	The fuzzy system is applied to all the solutions of the current population. Each solution is given a fuzzy score, and the tour with the highest score is used as the new reference tour.
\end{enumerate}
	
\subsubsection{Illustrative example}
Consider a stagnation point at which the fuzzy controller is invoked. Suppose we have five candidate tours $\{s_1,\dots,s_5\}$.
	
\textit{Step 1 -- Fuzzification.} The five tours with normalized objectives $(\text{RD},\text{CD})$ are evaluated using the membership functions of Figure~\ref{fig:membership_function}. That is, the crisp inputs are mapped onto the related linguistic terms (see Table~\ref{tab:linguistic_var}). The resulting membership degrees are reported in Table~\ref{tab:fuzz}.
	
\begin{table}[htbp]
	\centering
	\caption{Fuzzification of the five candidate solutions.}
	\label{tab:fuzz}
	\begin{tabular}{@{}l|ccc|ccc@{}}
		\toprule
		Solution & RD & $\mu_{\text{Small}}^{\text{RD}}$ & $\mu_{\text{Long}}^{\text{RD}}$ & CD & $\mu_{\text{Low}}^{\text{CD}}$ & $\mu_{\text{High}}^{\text{CD}}$ \\
		\midrule
		$s_1$ & 0.73 & 0.000 & 0.943 & 0.14 & 1.000 & 0.000 \\
		$s_2$ & 0.58 & 0.057 & 0.514 & 0.10 & 1.000 & 0.000 \\
		$s_3$ & 0.90 & 0.000 & 1.000 & 0.20 & 1.000 & 0.000 \\
		$s_4$ & 0.42 & 0.514 & 0.057 & 0.56 & 0.114 & 0.457 \\
		$s_5$ & 0.80 & 0.000 & 1.000 & 0.09 & 1.000 & 0.000 \\
		\bottomrule
	\end{tabular}
\end{table}
	
\textit{Step 2 -- Rule evaluation and aggregation.} Using Mamdani min--max inference, the four rules of Section~\ref{Fuzzy_syst} yield the firing strengths of Table~\ref{tab:rules}. Output terms pointing to the same consequent are aggregated by the maximum operator.
	
\begin{table}[htbp]
\centering
\caption{Rule activation and aggregated output levels.}
\label{tab:rules}
\begin{tabular}{@{}lccccccc@{}}
	\toprule
	\multirow{2}{*}{Chr.}  & \multicolumn{4}{c}{Rule activation (min)} & \multicolumn{3}{c}{Subregions (max)} \\
	\cmidrule(lr){2-5}\cmidrule(lr){6-8}
	 & R1 & R2 & R3 & R4 & Low & Medium & High \\
	\midrule
	$s_1$ & 0.000 & 0.943 & 0.000 & 0.000 & 0.000 & 0.943 & 0.000 \\
	$s_2$ & 0.057 & 0.514 & 0.000 & 0.000 & 0.000 & 0.514 & 0.057 \\
	$s_3$ & 0.000 & 1.000 & 0.000 & 0.000 & 0.000 & 1.000 & 0.000 \\
	$s_4$ & 0.114 & 0.057 & 0.457 & 0.057 & 0.057 & 0.457 & 0.114 \\
	$s_5$ & 0.000 & 1.000 & 0.000 & 0.000 & 0.000 & 1.000 & 0.000 \\
	\bottomrule
\end{tabular}
\end{table}
	
Hence, for each solution, the fuzzy subregions are aggregated into a single fuzzy region defined by its characteristic vertices. For example, for $s_2$, the aggregated fuzzy region is represented by the vertices $(0,0)\rightarrow(0.300,0)\rightarrow(0.403,0.514)\rightarrow(0.597,0.514)\rightarrow(0.689,0.057)\rightarrow(1.000,0.057)\rightarrow(1.000,0)$.
This polygon is subsequently used in the defuzzification step.
	
\textit{Step 3 -- Defuzzification and selection.} The crisp selection score is obtained by computing the $x$-coordinate of the centroid of the aggregated fuzzy region. The resulting scores are reported in Table~\ref{tab:score}. Consequently, $s_2$ is selected as the new reference tour, as it achieves the highest score, indicating that it best balances route distance and customer dissatisfaction.
	
\begin{table}[htbp]
\centering
\caption{Final ranking of the candidate reference tours.}
\label{tab:score}
\begin{tabular}{@{}clc@{}}
\toprule
Rank & Solution & Crisp selection score\\
\midrule
1 & $s_2$ & \textbf{0.5356}  \\
2 & $s_4$ & 0.5319  \\
3 & $s_1$ & 0.5000  \\
4 & $s_3$ & 0.5000 \\
5 & $s_5$ & 0.5000  \\
\bottomrule
\end{tabular}
\end{table}
	
\begin{figure}[htbp]
	\centering
	\includegraphics[width=0.85\textwidth]{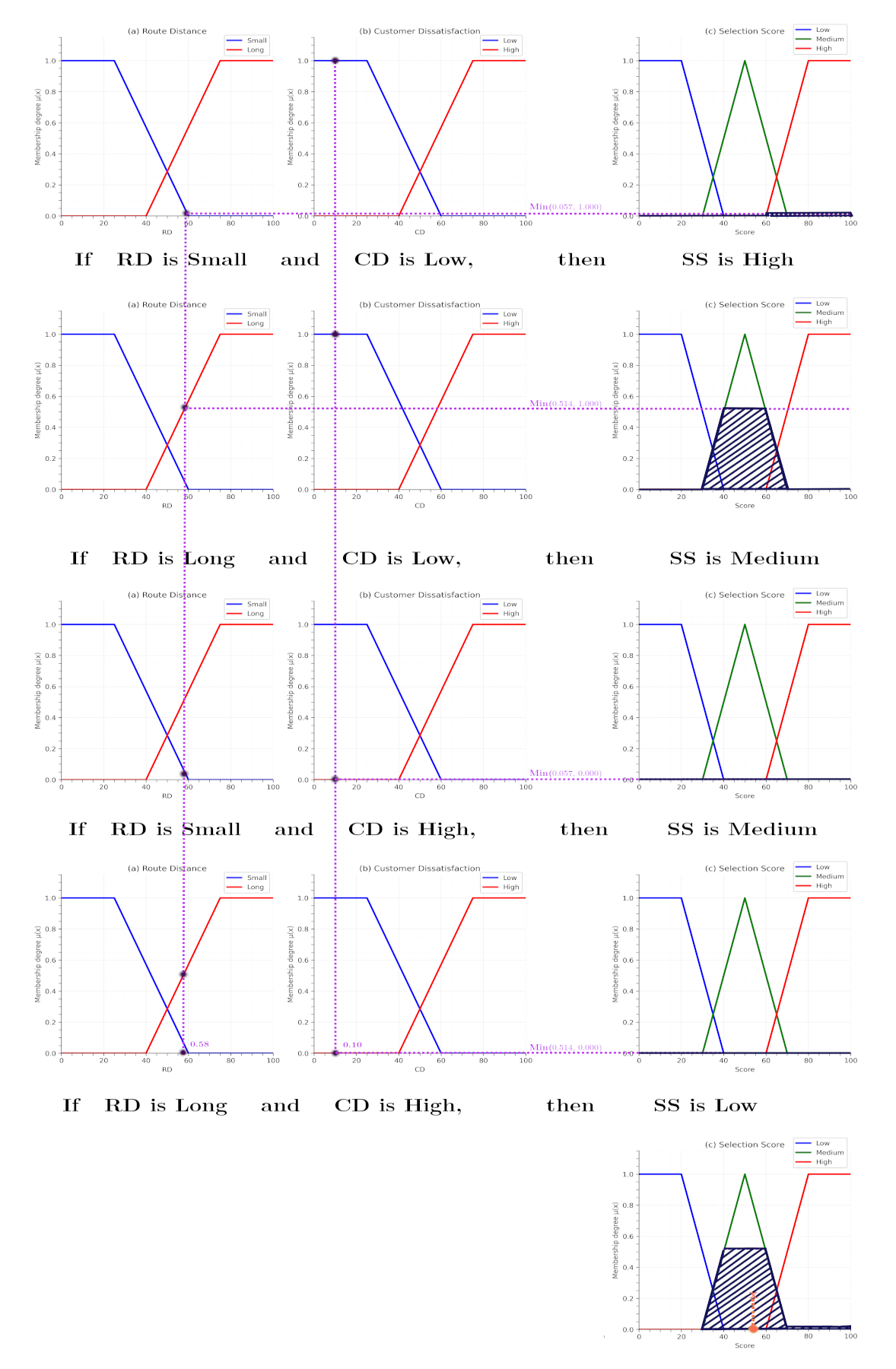}
	\caption{Inference computations for the second candidate solution.}
	\label{fig:fuzzyInfer}
\end{figure}

Figure  \ref{fig:fuzzyInfer} provides an overview of the three-step fuzzy inference process for candidate solution $s_2$

\subsubsection{Integration in GA process}
\begin{figure}[htbp]
	\centering
	\includegraphics[width=0.85\textwidth]{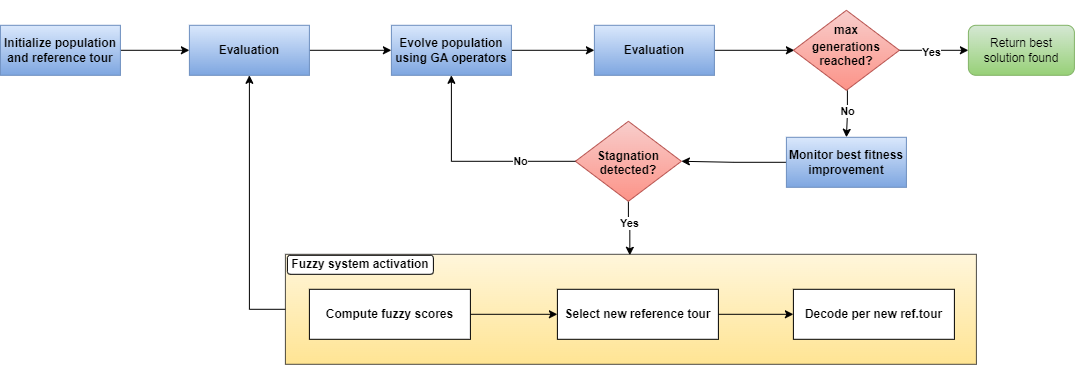}
	\caption{Flowchart of the proposed FL-NSE Genetic Algorithm.}
	\label{fig:Flowchart}
\end{figure}

The proposed approach operates by initially applying the classical NSE-based GA using a randomly generated reference tour. The evolutionary process monitors the improvement of the best fitness value at each generation. When stagnation is detected, that is, no significant improvement is observed over a predefined number of consecutive generations, the fuzzy decision process is activated. This mechanism evaluates all candidate solutions in the current population using \textit{RD} and \textit{CD} criteria defined in Table \ref{tab:linguistic_var}, and subsequently selects a high-quality reference tour to guide the continuing search. The detailed algorithmic workflow is represented in Fig.\ref{fig:Flowchart}.

\section{Experimental setup}
\label{sec:experiments}
	
\subsection{Benchmark instances}
	
We selected 13 standard instances from the TSPLIB library to evaluate the proposed approach. These instances range from 51 to 200 cities, covering a spectrum of problem scales that allows us to observe how the fuzzy adaptation behaves as the size grows. Table \ref{tab:instances} lists the test instances along with their best-known single-objective TSP distance performances. These values serve strictly as reference benchmarks rather than direct optimization targets. Because our formulation incorporates { position-based service requirements} via a weighted-sum objective, minimizing distance alone is no longer the sole goal, and a distance-optimal tour does not necessarily yield an optimal overall performance under the combined objective function.
	
\begin{table}[htbp]
	\centering
	\caption{TSPLIB problem instances used in experiments.}
	\label{tab:instances}
	\begin{tabular}{@{}clrr@{}}
	\toprule
	\textbf{No.} & \textbf{Instance} & \textbf{$\#$ of clients} & \textbf{Best known} \\
	\midrule
	1 & eil51 & 51 & 426 \\
	2 & berlin52 & 52 & 7,542 \\
	3 & st70 & 70 & 675 \\
	4 & eil76 & 76 & 538 \\
	5 & pr76 & 76 & 108,159 \\
	6 & kroA100 & 100 & 21,282 \\
	7 & kroB100 & 100 & 22,141 \\
	8 & kroC100 & 100 & 20,749 \\
	9 & eil101 & 101 & 629 \\
	10 & lin105 & 105 & 14,379 \\
	11 & pr144 & 144 & 58,537 \\
	12 & kroA150 & 150 & 26,524 \\
	13 & kroA200 & 200 & 29,368 \\
	\bottomrule
	\end{tabular}
\end{table}
	
{\subsection{Generating position-priority data}}
	
Since TSPLIB benchmarks natively lack preference attributes, we augmented the original coordinate-based datasets by generating synthetic {position-priority} data. For each city $i$, a  maximum visit rank $p_i$ is drawn uniformly from $\{1, \ldots, n\}$. This means a customer might be { associated with} an early visit rank (low $p_i$) or tolerate a later position (high $p_i$), with no geographic pattern imposed (That is, preferences are independent to city locations). We chose uniform random generation for a specific reason: it eliminates any hidden structure that might accidentally help or hurt either algorithm, giving us a clean test of whether the fuzzy mechanism genuinely improves search capability.
		
\subsection{Algorithm parameters}
	
We tuned parameters experimentally by testing all combinations within the following ranges:
	
\begin{itemize}
    \item Population size: $\{200, 800, 1000, 1500\}$;
    \item Number of iterations: $\{500, 1000, 1500, 2500\}$;
    \item Mutation probability: $\{0.01, 0.03, 0.05, 0.10\}$.
\end{itemize}
		
We set the crossover probability to $0.8$ and the elite proportion to $0.1$, based on preliminary parameter experiments and commonly used settings in genetic-algorithm implementations for routing problems.To account for varying search complexities, we determined the stagnation threshold experimentally for each problem instance.	
	
The bi-objective function weights remain fixed at $w_1 = w_2 = 0.5$ for all instances. Before applying these weights, each objective is normalized by a corresponding upper bound to prevent	differences in numerical scale from distorting the aggregation. Equal weighting is therefore applied to the normalized objectives, reflecting our intention to treat travel distance and customer satisfaction as equally important without imposing an a priori preference for either criterion. Consequently, the algorithm must genuinely balance both objectives rather than favouring one at the expense of the other.
	
We evaluated the algorithm's performance across four metrics: total travel distance, customer dissatisfaction, combined fitness value, and computational time. To account for its stochastic nature, we executed 30 independent runs per problem instance and recorded the mean values, ensuring a statistically robust assessment of the algorithm's behaviour.
	
\subsection{Computational environment}
	
All experiments were conducted on a single workstation powered by an Intel Core i3 processor (2.0 GHz) and 4 GB of RAM. The formulation of Section \ref{linFormul} was additionally coded in GNU MathProg and solved with GLPK to obtain exact baseline solutions, using a 4-hour time limit per instance. The resulting empirical evaluations are presented in the following section.
	
\section{Experiment results}
\label{sec:results}
Table \ref{tab:results} reports the comparative performance analysis of the proposed fuzzy enhanced NSE (FL-NSE) vs. classical NSE on 13 TSPLIB benchmarks of varying difficulty with number of cities in range from 51 to 200. Each comparison was performed through $30$ independent experiments taking into account stochastic nature of the algorithm. The table also includes the best incumbent returned by GLPK within a 4-hour time limit, where '--' indicates that no feasible incumbent was found within the allotted computation time.
	
\begin{sidewaystable}[htbp]
	\centering
	\caption{Results}
	\label{tab:results}
	\resizebox{\textwidth}{!}{%
	\begin{tabular}{cllllllllllllllllllc}
	\toprule
	\multirow{2}{*}{\textbf{Class}} & \multirow{2}{*}{\textbf{Instance}} & \multirow{2}{*}{\textbf{Method}} & \multicolumn{5}{c}{\textbf{Fitness}} & \multicolumn{5}{c}{\textbf{Distance}} & \multicolumn{5}{c}{\textbf{Dissatisfaction}} & \multirow{2}{*}{\textbf{Time (s)}} &
	\multirow{2}{*}{\textbf{GLPK}} \\
	\cmidrule(lr){4-8} \cmidrule(lr){9-13} \cmidrule(lr){14-18}
	 &  &  & \textbf{Min} & \textbf{Avg} & \textbf{Max} & \textbf{Med.} & \textbf{Std} & \textbf{Min} & \textbf{Avg} & \textbf{Max} & \textbf{Med.} & \textbf{Std} & \textbf{Min} & \textbf{Avg} & \textbf{Max} & \textbf{Med.} & \textbf{Std} &  \\
	\midrule
	\multirow{11}{*}{Small} 
	 & \multirow{2}{*}{eil51} & \cellcolor{gray!20}FL-NSE & \cellcolor{gray!20}\textbf{0.234287} & \cellcolor{gray!20}\textbf{0.269317} & \cellcolor{gray!20}\textbf{0.300863} & \cellcolor{gray!20}\textbf{0.26696} & \cellcolor{gray!20}\textbf{0.014015} & \cellcolor{gray!20}\textbf{739.2} & \cellcolor{gray!20}\textbf{858.34} & \cellcolor{gray!20}\textbf{986.27} & \cellcolor{gray!20}\textbf{858.18} & \cellcolor{gray!20}\textbf{61.24} & \cellcolor{gray!20}\textbf{138} & \cellcolor{gray!20}\textbf{199.17} & \cellcolor{gray!20}\textbf{244} & \cellcolor{gray!20}\textbf{199.5} & \cellcolor{gray!20}\textbf{22.71} & \cellcolor{gray!20}70.08 &  \multirow{2}{*}{0.283176}  \\
	 &  & NSE & 0.248812 & 0.28293 & 0.31158 & 0.282858 & 0.01659 & 774.09 & 878.76 & 1002.51 & 886.04 & 62.74 & 170 & 221.3 & 282 & 224 & 29.18 & 45.21 &   \\
	\cmidrule(lr){2-20}
	 & \multirow{2}{*}{berlin52} & \cellcolor{gray!20}FL-NSE & \cellcolor{gray!20}\textbf{0.25868} & \cellcolor{gray!20}\textbf{0.294321} & \cellcolor{gray!20}\textbf{0.323625} & \cellcolor{gray!20}\textbf{0.294797} & \cellcolor{gray!20}0.016121 & \cellcolor{gray!20}\textbf{13375.21} & \cellcolor{gray!20}\textbf{15338.37} & \cellcolor{gray!20}\textbf{17418.03} & \cellcolor{gray!20}\textbf{15323.8} & \cellcolor{gray!20}1010.11 & \cellcolor{gray!20}\textbf{181} & \cellcolor{gray!20}\textbf{232.4} & \cellcolor{gray!20}\textbf{295} & \cellcolor{gray!20}\textbf{230.5} & \cellcolor{gray!20}\textbf{28.77} & \cellcolor{gray!20}75.38 &  \multirow{2}{*}{0.305344} \\
	 &  & NSE & 0.276412 & 0.309878 & 0.339123 & 0.309032 & \textbf{0.015052} & 14064.82 & 16024.78 & 17818.59 & 15893.83 & \textbf{959.7} & 187 & 248.07 & 332 & 245.5 & 29.68 & 45.05 &  \\
	\cmidrule(lr){2-20}
	 & \multirow{2}{*}{st70} & \cellcolor{gray!20}FL-NSE & \cellcolor{gray!20}\textbf{0.248766} & \cellcolor{gray!20}\textbf{0.27216} & \cellcolor{gray!20}\textbf{0.298308} & \cellcolor{gray!20}\textbf{0.272221} & \cellcolor{gray!20}\textbf{0.01182} & \cellcolor{gray!20}1611.59 & \cellcolor{gray!20}\textbf{1803.77} & \cellcolor{gray!20}\textbf{2120.37} & \cellcolor{gray!20}\textbf{1783.42} & \cellcolor{gray!20}\textbf{112.24} & \cellcolor{gray!20}\textbf{426} & \cellcolor{gray!20}\textbf{508.57} & \cellcolor{gray!20}\textbf{639} & \cellcolor{gray!20}\textbf{505.5} & \cellcolor{gray!20}55.91 & \cellcolor{gray!20}252.33 & \multirow{2}{*}{0.261171} \\
	 &  & NSE & 0.253256 & 0.283462 & 0.31578 & 0.285908 & 0.017667 & \textbf{1528.67} & 1822.89 & 2126.06 & 1818.77 & 144.77 & 472 & 556.7 & 648 & 568.5 & \textbf{48.88} & 117.34 &  \\
	\cmidrule(lr){2-20}
	 & \multirow{2}{*}{eil76} & \cellcolor{gray!20}FL-NSE & \cellcolor{gray!20}0.261437 & \cellcolor{gray!20}\textbf{0.283642} & \cellcolor{gray!20}\textbf{0.306862} & \cellcolor{gray!20}\textbf{0.284082} & \cellcolor{gray!20}\textbf{0.011308} & \cellcolor{gray!20}1127.97 & \cellcolor{gray!20}\textbf{1285.52} & \cellcolor{gray!20}\textbf{1453.91} & \cellcolor{gray!20}1283.76 & \cellcolor{gray!20}\textbf{87.65} & \cellcolor{gray!20}\textbf{534} & \cellcolor{gray!20}\textbf{619.93} & \cellcolor{gray!20}\textbf{719} & \cellcolor{gray!20}\textbf{629.5} & \cellcolor{gray!20}\textbf{48.07} & \cellcolor{gray!20}280.78 & \multirow{2}{*}{0.282543}  \\
	 &  & NSE & \textbf{0.252838} & 0.289322 & 0.324957 & 0.288061 & 0.017514 & \textbf{1069.95} & 1286.16 & 1461.48 & \textbf{1279.27} & 99.42 & 571 & 669.67 & 773 & 664.5 & 53.2 & 144.24 &  \\
	\cmidrule(lr){2-20}
	 & \multirow{2}{*}{pr76} & \cellcolor{gray!20}FL-NSE & \cellcolor{gray!20}\textbf{0.217188} & \cellcolor{gray!20}\textbf{0.254838} & \cellcolor{gray!20}\textbf{0.285303} & \cellcolor{gray!20}\textbf{0.253432} & \cellcolor{gray!20}0.015492 & \cellcolor{gray!20}\textbf{238197.58} & \cellcolor{gray!20}\textbf{278499.48} & \cellcolor{gray!20}\textbf{309437.5} & \cellcolor{gray!20}\textbf{279559.96} & \cellcolor{gray!20}20832.41 & \cellcolor{gray!20}\textbf{294} & \cellcolor{gray!20}\textbf{428.07} & \cellcolor{gray!20}\textbf{513} & \cellcolor{gray!20}\textbf{442.5} & \cellcolor{gray!20}54.94 & \cellcolor{gray!20}272.89 & \multirow{2}{*}{0.267799} \\
	 &  & NSE & 0.245146 & 0.273503 & 0.299608 & 0.270854 & \textbf{0.014177} & 246319.18 & 290981.23 & 322629.82 & 293474.2 & \textbf{19322.16} & 383 & 485.93 & 589 & 485 & \textbf{49.25} & 145.39 &  \\
	\midrule
	\multirow{11}{*}{Medium} 
	 & \multirow{2}{*}{kroA100} & \cellcolor{gray!20}FL-NSE & \cellcolor{gray!20}\textbf{0.233938} & \cellcolor{gray!20}\textbf{0.272177} & \cellcolor{gray!20}0.312997 & \cellcolor{gray!20}\textbf{0.27207} & \cellcolor{gray!20}\textbf{0.015075} & \cellcolor{gray!20}\textbf{67568.32} & \cellcolor{gray!20}\textbf{79614.93} & \cellcolor{gray!20}\textbf{91468.45} & \cellcolor{gray!20}\textbf{78191.49} & \cellcolor{gray!20}\textbf{5486.86} & \cellcolor{gray!20}\textbf{1025} & \cellcolor{gray!20}\textbf{1171.47} & \cellcolor{gray!20}1460 & \cellcolor{gray!20}\textbf{1147.5} & \cellcolor{gray!20}101.9 & \cellcolor{gray!20}425.46 & \multirow{2}{*}{0.260242} \\
	 &  & NSE & 0.255683 & 0.283519 & \textbf{0.306969} & 0.285293 & 0.015391 & 70179.05 & 81326.45 & 95564.33 & 81631.49 & 5590.89 & 1121 & 1254 & \textbf{1424} & 1258.5 & \textbf{79.67} & 330.78 &  \\
	\cmidrule(lr){2-20}
	 & \multirow{2}{*}{KroB100} & \cellcolor{gray!20}FL-NSE & \cellcolor{gray!20}0.24033 & \cellcolor{gray!20}\textbf{0.263147} & \cellcolor{gray!20}\textbf{0.288118} & \cellcolor{gray!20}\textbf{0.263988} & \cellcolor{gray!20}\textbf{0.01311} & \cellcolor{gray!20}67066.1 & \cellcolor{gray!20}\textbf{78113.7} & \cellcolor{gray!20}\textbf{88817.65} & \cellcolor{gray!20}\textbf{77440.98} & \cellcolor{gray!20}\textbf{4724.83} & \cellcolor{gray!20}\textbf{882} & \cellcolor{gray!20}\textbf{999.33} & \cellcolor{gray!20}\textbf{1164} & \cellcolor{gray!20}\textbf{1002} & \cellcolor{gray!20}\textbf{69.81} & \cellcolor{gray!20}427.5 & \multirow{2}{*}{--} \\
	 &  & NSE & \textbf{0.226128} & 0.272548 & 0.30408 & 0.27857 & 0.017314 & \textbf{63992.58} & 79232.59 & 89527.78 & 80702.03 & 6659.3 & 922 & 1068.8 & 1218 & 1070.5 & 75.05 & 339.02 &  \\
	\cmidrule(lr){2-20}
	 & \multirow{2}{*}{KroC100} & \cellcolor{gray!20}FL-NSE & \cellcolor{gray!20}\textbf{0.222298} & \cellcolor{gray!20}\textbf{0.247119} & \cellcolor{gray!20}\textbf{0.271965} & \cellcolor{gray!20}\textbf{0.244812} & \cellcolor{gray!20}\textbf{0.013092} & \cellcolor{gray!20}72433.12 & \cellcolor{gray!20}\textbf{79150.95} & \cellcolor{gray!20}\textbf{89046.89} & \cellcolor{gray!20}\textbf{78190.99} & \cellcolor{gray!20}\textbf{4690.38} & \cellcolor{gray!20}\textbf{671} & \cellcolor{gray!20}\textbf{834.33} & \cellcolor{gray!20}\textbf{1047} & \cellcolor{gray!20}\textbf{848} & \cellcolor{gray!20}105.19 & \cellcolor{gray!20}425.67 & \multirow{2}{*}{0.243862}  \\
	 &  & NSE & 0.224942 & 0.256719 & 0.29193 & 0.259508 & 0.016233 & \textbf{70069.13} & 79353.08 & 92946.25 & 79169.51 & 5541.42 & 690 & 920.23 & 1163 & 923.5 & \textbf{97.12} & 358.05 &  \\
	\cmidrule(lr){2-20}
	 & \multirow{2}{*}{eil101} & \cellcolor{gray!20}FL-NSE & \cellcolor{gray!20}\textbf{0.258797} & \cellcolor{gray!20}\textbf{0.284784} & \cellcolor{gray!20}\textbf{0.3168} & \cellcolor{gray!20}\textbf{0.286331} & \cellcolor{gray!20}\textbf{0.013249} & \cellcolor{gray!20}\textbf{1578.34} & \cellcolor{gray!20}\textbf{1767.14} & \cellcolor{gray!20}\textbf{1954.17} & \cellcolor{gray!20}\textbf{1767.65} & \cellcolor{gray!20}\textbf{88.27} & \cellcolor{gray!20}\textbf{867} & \cellcolor{gray!20}\textbf{1075.5} & \cellcolor{gray!20}\textbf{1298} & \cellcolor{gray!20}\textbf{1074.5} & \cellcolor{gray!20}\textbf{105.19} & \cellcolor{gray!20}552.15  & \multirow{2}{*}{0.254829} \\
	 &  & NSE & 0.266534 & 0.294183 & 0.32976 & 0.293924 & 0.014688 & 1595.81 & 1800.5 & 2029.77 & 1801.28 & 103.57 & 900 & 1137.77 & 1366 & 1123 & 106.58 & 385.51 &  \\
	\cmidrule(lr){2-20}
	 & \multirow{2}{*}{lin105} & \cellcolor{gray!20}FL-NSE & \cellcolor{gray!20}0.237249 & \cellcolor{gray!20}\textbf{0.262615} & \cellcolor{gray!20}\textbf{0.286841} & \cellcolor{gray!20}\textbf{0.26269} & \cellcolor{gray!20}\textbf{0.015117} & \cellcolor{gray!20}\textbf{48187.86} & \cellcolor{gray!20}\textbf{57363.14} & \cellcolor{gray!20}\textbf{64494.22} & \cellcolor{gray!20}\textbf{58273.14} & \cellcolor{gray!20}\textbf{4440.99} & \cellcolor{gray!20}\textbf{883} & \cellcolor{gray!20}\textbf{1069.83} & \cellcolor{gray!20}\textbf{1224} & \cellcolor{gray!20}\textbf{1073} & \cellcolor{gray!20}\textbf{99} & \cellcolor{gray!20}420.73  &  \multirow{2}{*}{0.245795} \\
	 &  & NSE & \textbf{0.233663} & 0.271498 & 0.307742 & 0.271934 & 0.016149 & 49316.15 & 58746.65 & 66716.23 & 58450.45 & 4545.39 & 933 & 1122.07 & 1361 & 1124 & 107.87 & 341.29  &    \\
	\midrule
	\multirow{7}{*}{Large} 
	 & \multirow{2}{*}{pr144} & \cellcolor{gray!20}FL-NSE & \cellcolor{gray!20}\textbf{0.265714} & \cellcolor{gray!20}\textbf{0.293158} & \cellcolor{gray!20}\textbf{0.338795} & \cellcolor{gray!20}\textbf{0.294429} & \cellcolor{gray!20}0.016894 & \cellcolor{gray!20}390120.39 & \cellcolor{gray!20}\textbf{433026.12} & \cellcolor{gray!20}\textbf{476288.38} & \cellcolor{gray!20}\textbf{433031.89} & \cellcolor{gray!20}\textbf{24804} & \cellcolor{gray!20}\textbf{1666} & \cellcolor{gray!20}\textbf{2119.9} & \cellcolor{gray!20}2728 & \cellcolor{gray!20}\textbf{2097} & \cellcolor{gray!20}224.44 & \cellcolor{gray!20}290.29  &  \multirow{2}{*}{--} \\
	 &  & NSE & 0.268054 & 0.302001 & 0.339316 & 0.303067 & \textbf{0.015784} & \textbf{375980.73} & 447509.65 & 511839.18 & 445113.82 & 25108.01 & 1763 & 2172.13 & \textbf{2430} & 2177.5 & \textbf{177.15} & 210.72  &   \\
	\cmidrule(lr){2-20}
	 & \multirow{2}{*}{kroA150} & \cellcolor{gray!20}FL-NSE & \cellcolor{gray!20}\textbf{0.248595} & \cellcolor{gray!20}\textbf{0.280391} & \cellcolor{gray!20}\textbf{0.304104} & \cellcolor{gray!20}\textbf{0.279866} & \cellcolor{gray!20}0.011722 & \cellcolor{gray!20}117545.15 & \cellcolor{gray!20}\textbf{128047.66} & \cellcolor{gray!20}\textbf{140998.6} & \cellcolor{gray!20}\textbf{127664.25} & \cellcolor{gray!20}\textbf{6468.78} & \cellcolor{gray!20}\textbf{1867} & \cellcolor{gray!20}\textbf{2423.6} & \cellcolor{gray!20}\textbf{2848} & \cellcolor{gray!20}\textbf{2455.5} & \cellcolor{gray!20}193.61 & \cellcolor{gray!20}890.05  &  \multirow{2}{*}{--} \\
	 &  & NSE & 0.27023 & 0.291987 & 0.317007 & 0.290739 & \textbf{0.011655} & \textbf{113401.8} & 132122.01 & 147751.04 & 133385.92 & 7329.81 & 2339 & 2559.5 & 2921 & 2581 & \textbf{142} & 999.99  &  \\
	\cmidrule(lr){2-20}
	 & \multirow{2}{*}{kroA200} & \cellcolor{gray!20}FL-NSE & \cellcolor{gray!20}\textbf{0.276154} & \cellcolor{gray!20}\textbf{0.318558} & \cellcolor{gray!20}\textbf{0.339808} & \cellcolor{gray!20}\textbf{0.319791} & \cellcolor{gray!20}0.012381 & \cellcolor{gray!20}\textbf{155942.06} & \cellcolor{gray!20}\textbf{186839.94} & \cellcolor{gray!20}\textbf{204711.45} & \cellcolor{gray!20}\textbf{188000.94} & \cellcolor{gray!20}9995.35 & \cellcolor{gray!20}\textbf{5053} & \cellcolor{gray!20}\textbf{5562.1} & \cellcolor{gray!20}\textbf{6090} & \cellcolor{gray!20}\textbf{5601} & \cellcolor{gray!20}301.2 & \cellcolor{gray!20}1010.23 & \multirow{2}{*}{--}  \\
	 &  & NSE & 0.313142 & 0.333565 & 0.351441 & 0.333501 & \textbf{0.008208} & 182712.8 & 198375.33 & 210321.52 & 198834.91 & \textbf{5645.27} & 5352 & 5710.6 & 6199 & 5698 & \textbf{209.23} & 1135.33  &   \\
	\bottomrule
	\end{tabular}
}
\end{sidewaystable}

To statistically assess the impact of the fuzzy logic mechanism, a paired Wilcoxon signed-rank test was performed to compare standard NSE and FL-NSE. The test was conducted using the \texttt{wilcox.test()} function from R's built-in \texttt{stats} package. Pairwise differences were computed as $d_i = \text{NSE}_i - \text{FL-NSE}_i$, where each difference represents the performance gap between the two variants based on the minimum fitness values (reported in the fourth column of Table~\ref{tab:results}) obtained from the $i$th paired run.	
The statistical test revealed a significant advantage in favor of FL-NSE ($V = 75$, $p = 0.03979$). Because $p < 0.05$, the null hypothesis of equal median fitness distributions is rejected in favor of a true location shift. Descriptive analysis of the pairwise differences confirms that FL-NSE consistently achieves lower fitness values, with a median difference ($\text{NSE} - \text{FL-NSE}$) of $+0.007737$ units and a mean difference of $+0.010108$ units. Furthermore, the first quartile ($+0.002340$) demonstrates that FL-NSE outperformed standard NSE in over $75\%$ of the benchmark runs, reaching maximum cost reductions up to $+0.036988$. These empirical results demonstrate that incorporating the fuzzy engine significantly enhances the algorithm's search capability and convergence quality.
	
\begin{figure}[h!]
	\centering
	\includegraphics[width=0.95\textwidth]{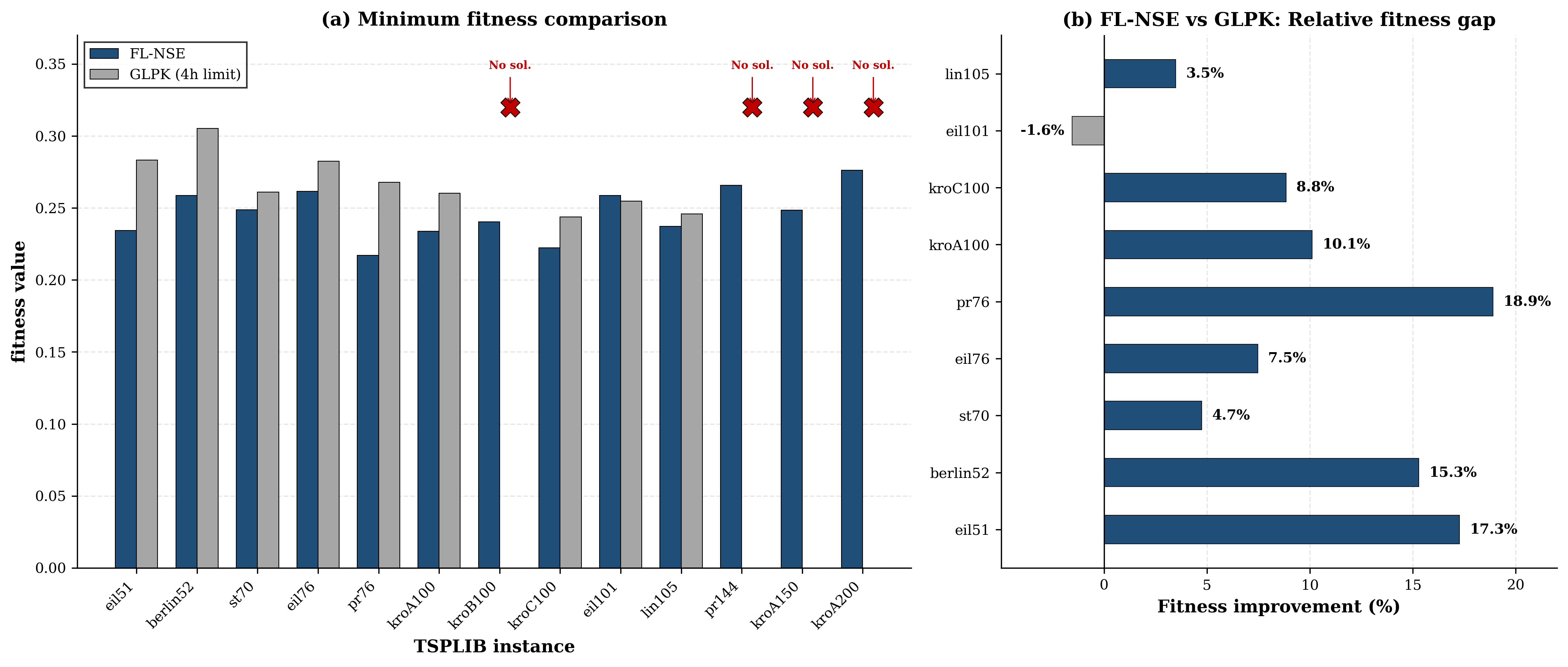}
	\caption{(a) Minimum fitness comparison between FL-NSE and GLPK (4h limit). Red cross indicates instances where GLPK found no feasible solution. 
	(b) Relative fitness gap on instances solved by GLPK.}
	\label{fig:FLNSE_GLPK}
\end{figure}

To assess the absolute quality of the obtained solutions, we solved the origin-independent linear formulation of Section~\ref{linFormul} with GLPK under a 4-hour time limit. The results expose a clear scalability gap. While GLPK successfully returned feasible incumbents for all small and medium-sized instances, it completely failed on every large instance (\texttt{kroB100}, \texttt{pr144}, \texttt{kroA150}, and \texttt{kroA200}), returning no feasible solution whatsoever within the 4-hour horizon. Strikingly, FL-NSE not only finds feasible tours for these very same instances but delivers high-quality solutions across all 13 benchmarks in a fraction of that time. Among the 9 instances where GLPK did succeed, FL-NSE achieves a lower minimum fitness on 8 cases, with relative fitness improvements ranging from +3.48\% (\texttt{lin105}) to +18.90\% (\texttt{pr76}, as illustrated in Figure~\ref{fig:FLNSE_GLPK}. Only on \texttt{eil101} does GLPK achieve a marginally better fitness ($-1.56\%$ relative to FL-NSE). These findings demonstrate that the proposed fuzzy-enhanced metaheuristic does not merely approximate exact solutions, it reliably solves problem instances on which an off-the-shelf ILP solver failed to find a feasible incumbent despite a 4-hour computation time limit.
	
\begin{figure}[h!]
	\centering
	\includegraphics[width=0.8\textwidth]{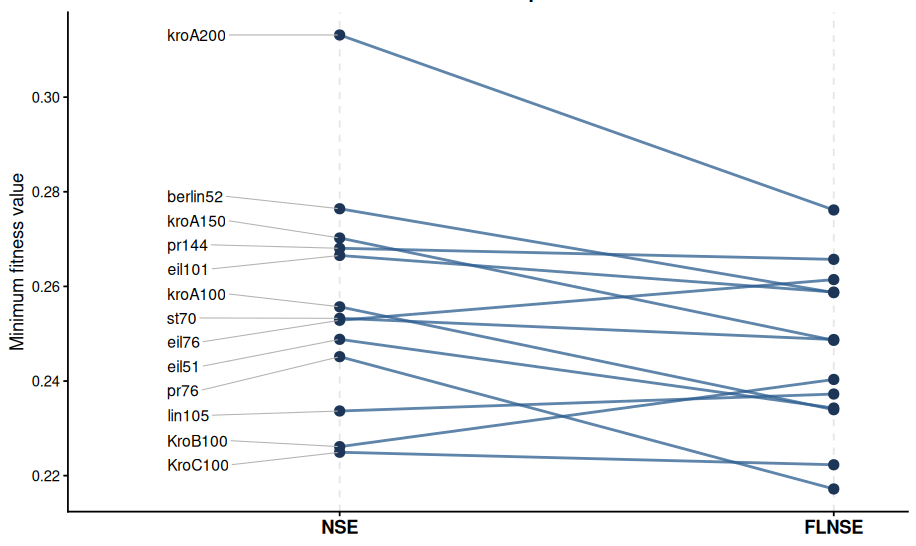}
	\caption{Instance-wise performance comparison.}
	\label{fig:slopegraph}
\end{figure}

To visually demonstrate the performance superiority of FL-NSE over standard NSE, Figure~\ref{fig:slopegraph} presents a slopegraph of the pairwise benchmark comparisons. As illustrated, the vast majority of connecting lines exhibit a downward slope from left to right, confirming the consistent cost reductions achieved by FL-NSE across individual instances. 
	
\begin{figure}[h!]
	\centering
	\includegraphics[width=0.95\textwidth]{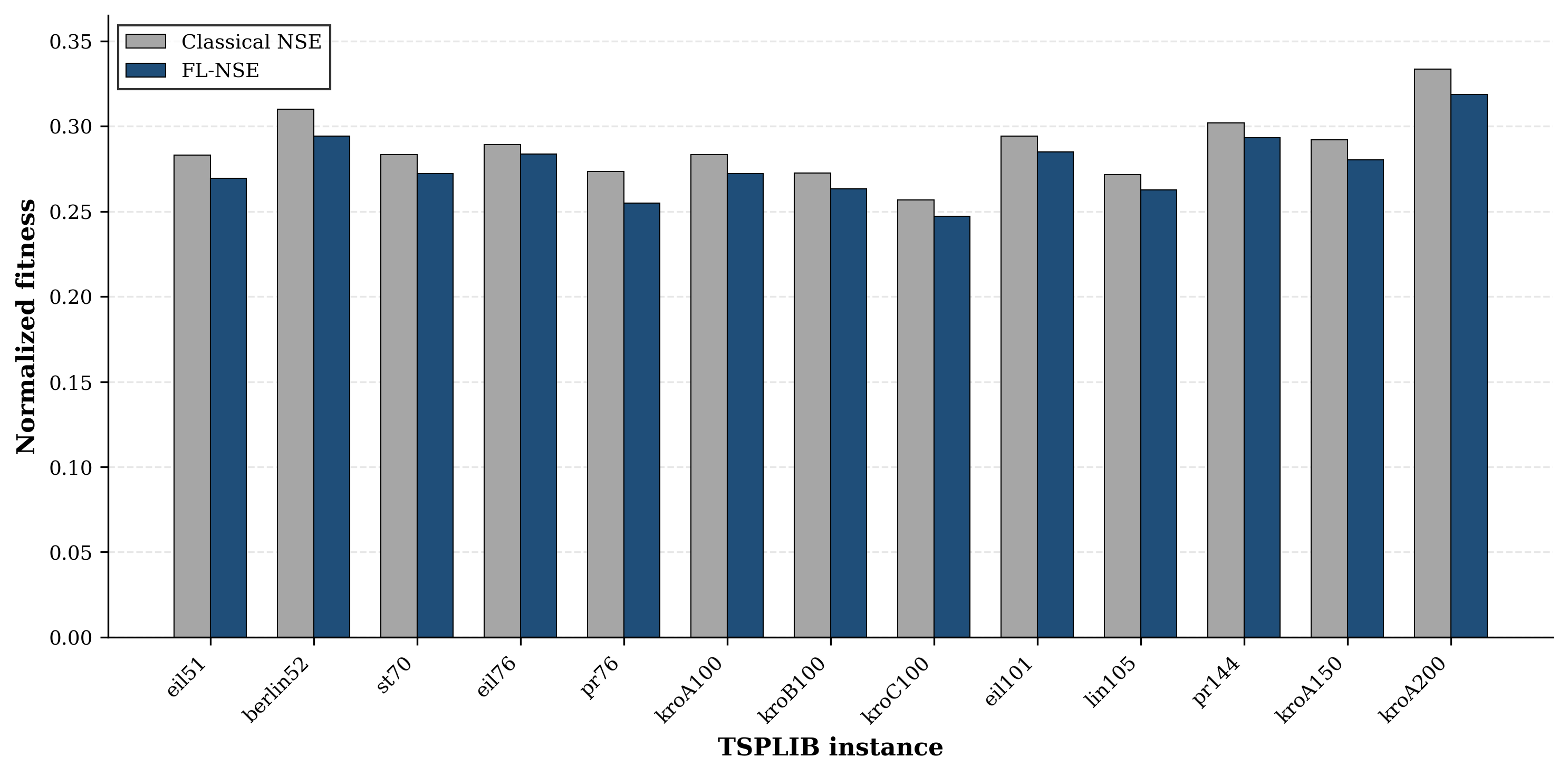}
	\caption{Comparative average fitness values across TSPLIB instances.}
	\label{fig:fitness_comparison}
\end{figure}

\begin{figure}[h!]
	\centering
	\includegraphics[width=0.95\textwidth]{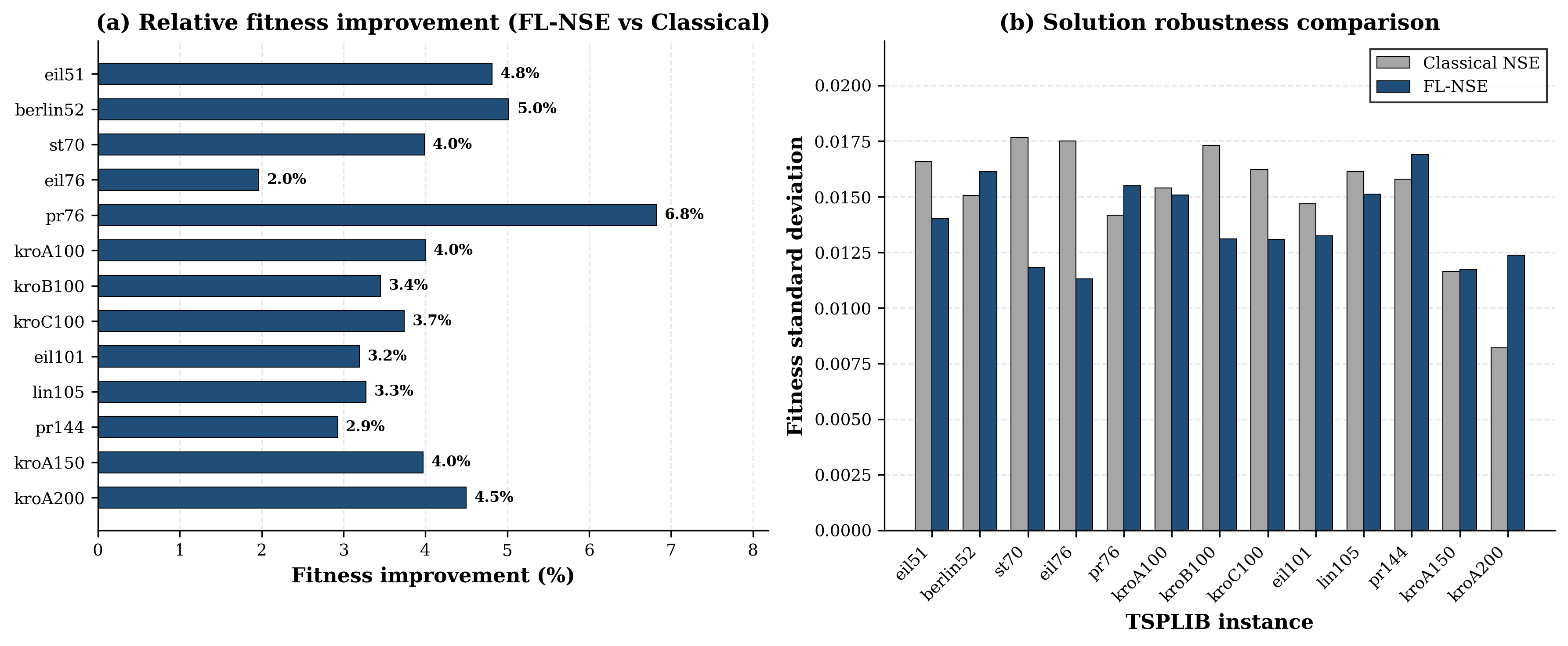}
	\caption{(a) Relative fitness improvement by instance; (b) Standard 
	deviation comparison.}
	\label{fig:improvement}
\end{figure}

Furthermore, the fuzzy-enhanced approach achieves lower average fitness values across all instances(see Fig.\ref{fig:fitness_comparison}), with relative improvements ranging from $2.0\%$ (\texttt{eil76}) to $6.8\%$ (\texttt{pr76}). This uniform superiority indicates that the dynamic reference tour update effectively guides the search toward better compromises between travel distance and customer dissatisfaction, regardless of problem size or structure. A closer examination of extreme values reveals that FL-NSE not only improves mean performance but also reduces worst-case scenarios; for instance, on \texttt{lin105}, the maximum fitness decreases from $0.3077$ to $0.2868$, corresponding to a $6.8\%$ reduction. While the fuzzy mechanism enhances overall robustness, a nuanced pattern emerges on specific instances; for example, on \texttt{eil76}, the best found solution with FL ($0.2614$) is marginally outperformed by that without FL ($0.2528$), suggesting that the dynamic update may occasionally overlook high-quality regions reached by the static encoding. The standard deviation of fitness values decreases on $8$ out of $13$ instances, particularly for medium scale problems such as \texttt{kroB100} and \texttt{kroC100}. However, on some large-scale instances (\texttt{kroA150} and \texttt{kroA200}), the standard deviation increases from $0.01166$ to $0.01172$ and from $0.0082$ to $0.0124$, respectively. These improvements are  quantified in Fig.\ref{fig:improvement}(a), while the enhanced  robustness is confirmed by the reduced standard deviation depicted in Fig.\ref{fig:improvement}(b)
	
\begin{figure}[h!]
	\centering
	\includegraphics[width=0.95\textwidth]{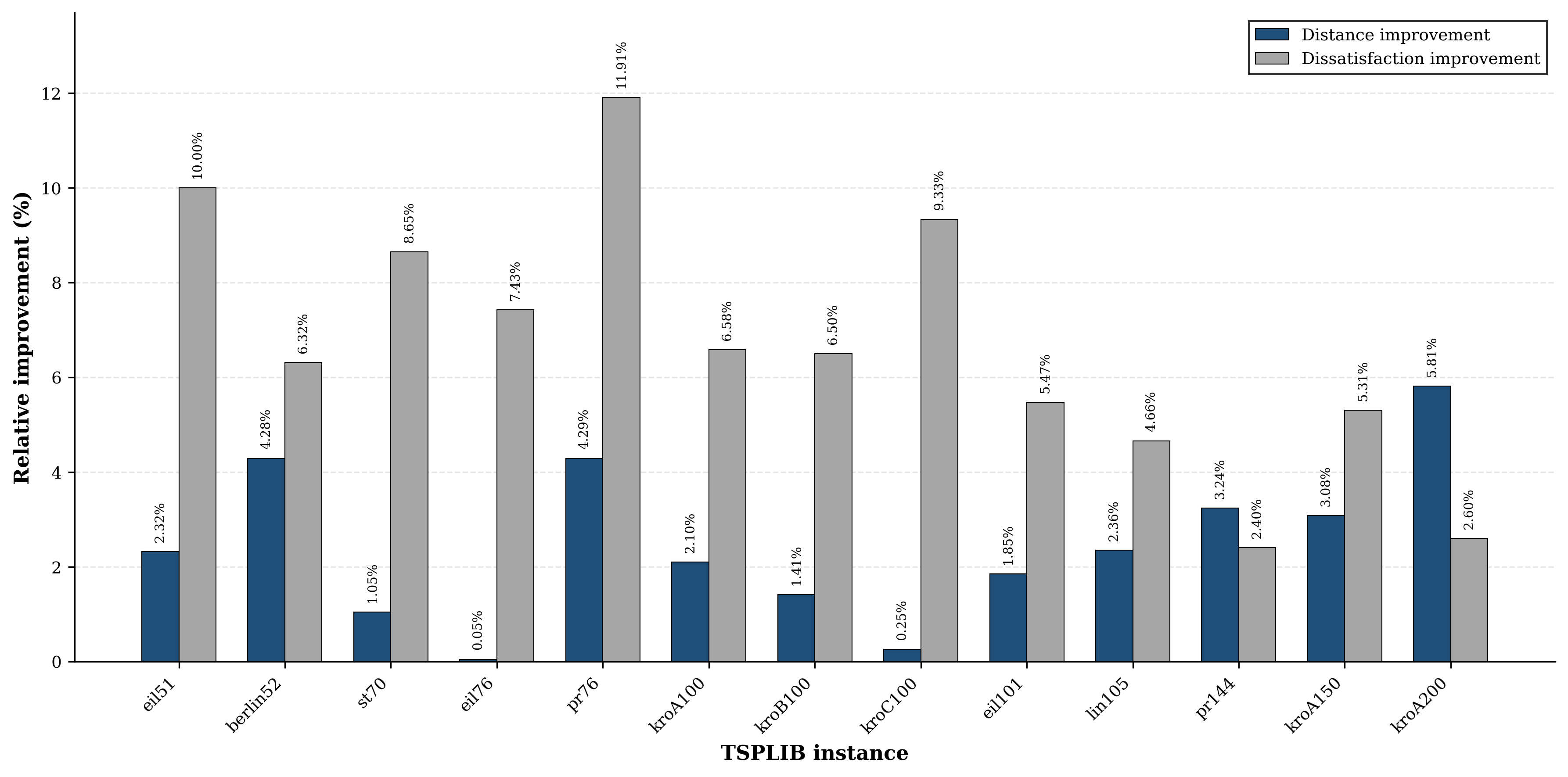}
	\caption{Relative improvement of FL-NSE over classical NSE for each 
	objective component.}
	\label{fig:dist_dissat}
\end{figure}

Contrary to the common expectation that improving service quality necessarily degrades routing efficiency, the results demonstrate that FL-NSE frequently improves both objectives simultaneously (see Fig.\ref{fig:dist_dissat}). For example, on \texttt{kroA200}, the average travel distance decreases from $198{,}375$ to $186{,}840$, while dissatisfaction is reduced from $5{,}710$ to $5{,}562$. Similarly, on \texttt{pr76}, the dissatisfaction improvement reaches $11.91\%$, the highest across all instances, demonstrating the effectiveness of the fuzzy mechanism for problems with strong { position-priority} constraints.
	
\begin{figure}[h!]
	\centering
	\includegraphics[width=0.95\textwidth]{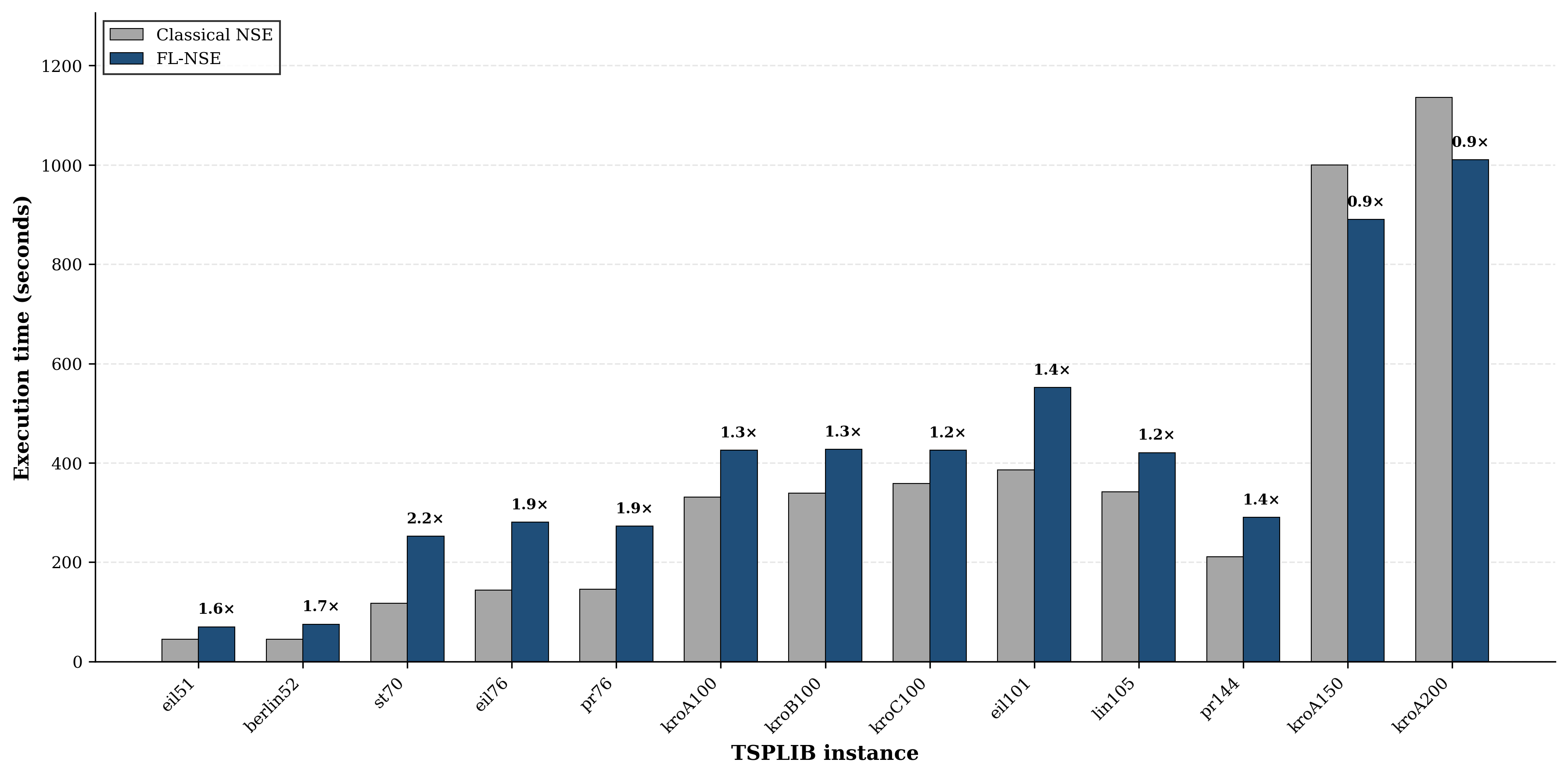}
	\caption{Computational time comparison}
	\label{fig:time}
\end{figure}

Regarding runtime, the proposed algorithm requires moderately more computation time compared to classical NSE on small and medium instances, with the ratio ranging from $1.2\times$ to $2.2\times$ (e.g.\texttt{st70}, \texttt{eil76}). However, on the largest instances (\texttt{kroA150} and \texttt{kroA200}), FL-NSE exhibits slightly lower execution times, with a runtime ratio of approximately $0.9\times$ (see Fig.~\ref{fig:time}). While this result is encouraging, the limited number of large-scale instances prevents any firm conclusions regarding the runtime behavior of the proposed approach.
	
\begin{figure}[h!]
	\centering
	\includegraphics[width=0.95\textwidth]{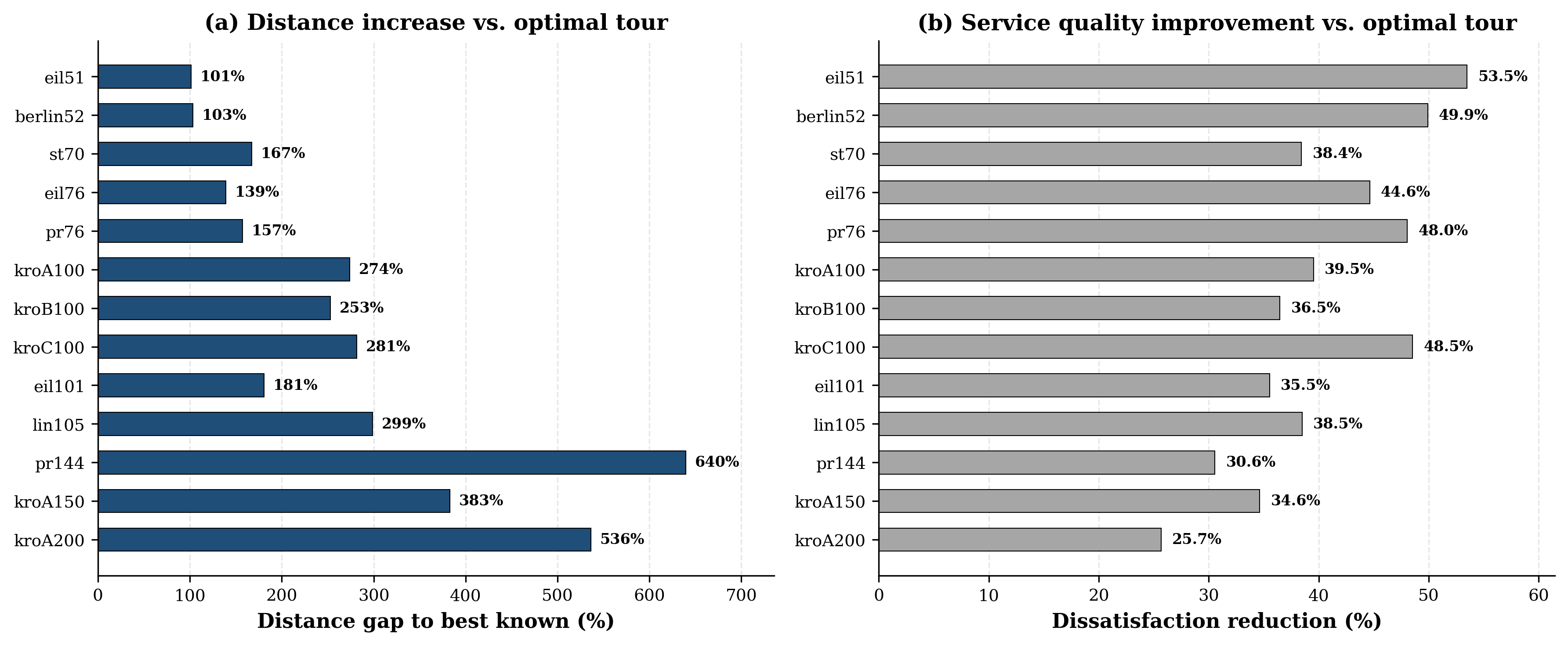}
	\caption{(a) Distance increase of FL-NSE relative to the best known 
	distance-optimal tour; (b) Corresponding dissatisfaction reduction achieved by FL-NSE compared to the optimal tour.}
	\label{fig:optimal}
\end{figure}

To further assess the effectiveness of the proposed approach, we compare the obtained solutions with the best known solutions of the classical TSP. Since a TSP tour is cyclic, the customer dissatisfaction associated with each distance-optimal tour is evaluated by fixing node 1 as the starting point. The reported values therefore correspond to this particular tour representation rather than the minimum possible dissatisfaction over all equivalent cyclic shifts. As reported in Table~\ref{tab:optimal}, although the distance-optimal tours achieve minimal travel costs, they incur substantial customer dissatisfaction.	
For instance, the optimal tour for \texttt{eil51} (distance 426) yields a dissatisfaction of 428, whereas FL-NSE achieves a dissatisfaction of 199 at the cost of doubling the travel distance. This pattern is consistent across all instances. Fig.\ref{fig:optimal} illustrates the magnitude of this trade-off: while the distance gap ranges from $+101\%$ to $+640\%$, the dissatisfaction reduction remains significant ($25.7\%$ to $53.5\%$). Notably, \texttt{pr144} exhibits the largest distance gap $640\%$ due to its complex spatial distribution, yet still achieves a substantial dissatisfaction reduction of $30.6\%$. This analysis highlights the limitations of using distance-optimal tours as benchmarks for overall solution quality in the bi-objective setting, since they are computed without considering {visit-rank constraints}.
\begin{table}[htbp]
\centering
\caption{Comparison with distance-optimal tours (Best known solutions).}
\label{tab:optimal}
\small
\begin{tabular}{@{}lcccccc@{}}
	\toprule
	\textbf{Instance} & \multicolumn{2}{c}{\textbf{Best known tour}} & \multicolumn{2}{c}{\textbf{FL-NSE (Avg.)}} & \textbf{Distance} & \textbf{Dissatisfaction} \\
	\cmidrule(lr){2-3} \cmidrule(lr){4-5}
	 & \textbf{Dist.} & \textbf{Dissat.} & \textbf{Dist.} & \textbf{Dissat.} & \textbf{Gap (\%)} & \textbf{Reduction (\%)} \\
	\midrule
	eil51 & 426 & 428 & 858 & 199 & +101.5 & 53.5 \\
	berlin52 & 7,542 & 464 & 15,338 & 232 & +103.4 & 49.9 \\
	st70 & 675 & 826 & 1,804 & 509 & +167.2 & 38.4 \\
	eil76 & 538 & 1,120 & 1,286 & 620 & +138.9 & 44.6 \\
	pr76 & 108,159 & 824 & 278,499 & 428 & +157.5 & 48.0 \\
	kroA100 & 21,282 & 1,937 & 79,615 & 1,171 & +274.1 & 39.5 \\
	kroB100 & 22,141 & 1,573 & 78,114 & 999 & +252.8 & 36.5 \\
	kroC100 & 20,749 & 1,621 & 79,151 & 834 & +281.5 & 48.5 \\
	eil101 & 629 & 1,668 & 1,767 & 1,076 & +180.9 & 35.5 \\
	lin105 & 14,379 & 1,739 & 57,363 & 1,070 & +298.9 & 38.5 \\
	pr144 & 58,537 & 3,053 & 433,026 & 2,120 & +639.7 & 30.6 \\
	kroA150 & 26,524 & 3,708 & 128,048 & 2,424 & +382.8 & 34.6 \\
	kroA200 & 29,368 & 7,482 & 186,840 & 5,562 & +536.2 & 25.7 \\
	\bottomrule
\end{tabular}
\end{table}
	
In summary, these collective findings underscore the critical role of adaptive reference tour management in Node Shift Encoding. In contrast to the static approach, where fixed reference tours progressively restrict neighbourhood exploration, our proposed fuzzy inference model systematically updates the search baseline using multi-criteria feedback. This continuous adaptation substantially mitigates search stagnation, ensuring sustained diversification and superior performance across the majority of benchmark instances.

\section{Discussion} 
\label{sec:discussion}
The question that motivated this work was straightforward: why does Node Shift Encoding, despite its theoretical properties, often plateau prematurely in practice? While several factors contribute to metaheuristic stagnation, our findings confirm that treating the reference tour as a static anchor plays an indisputably major role in restricting optimization performance. In theory, a single reference tour provides complete coverage of the search space, as a corresponding shift vector exists for every target permutation. Practically, however, navigating this space through a fixed baseline creates significant structural barriers, giving rise to localized stagnation and severe reachability constraints that can only be overcome through dynamic reference updating.
	
Across all 13 instances, FL-NSE improved upon classical NSE. This breadth is difficult to reconcile with incidental factors. It points, instead, to a structural tension: a good reference tour enables efficient local search but progressively narrows the reachable space; a poor one permits exploration but decodes to inferior solutions. Classical NSE makes this choice once, at initialization, with no mechanism to revisit it. The fuzzy controller we propose does exactly that — it revisits the choice when the search stalls.
	
The mechanism operates on a principle that differs subtly from the GA's own logic. The GA optimizes a scalarized sum, $w_1 D+w_2 C $ , which is vulnerable to a particular drift: solutions that balance poorly across objectives can still score well if their aggregate is favorable. Fuzzy logic, by contrast, evaluates distance and dissatisfaction as linguistically separate concepts , "short" versus "long," "low" versus "high",  and rewards solutions that achieve both. This is not parameter tuning; it is a different way of seeing the problem. This fuzzy-logic-driven epistemic difference, prioritizing bi-objective trade-off over a good weighted aggregate, explains why FL\-NSE consistently discovers solutions that dominate those accessible to classical NSE.
	
The results for instance \texttt{eil76} provide a compelling case for the necessity of our approach. While classical NSE produced a marginally better single best-found solution ($0.2528$ vs. $0.2614$), it exhibited significantly higher instability. In contrast, FL\-NSE demonstrated superior statistical reliability, achieving a lower mean fitness ($0.2836$ vs. $0.2893$) and a $35\%$ reduction in standard deviation ($0.0113$ vs. $0.0175$). This proves that FL\-NSE prioritizes consistent, high-quality performance over erratic, outlier successes , a distinction that defines operational robustness and ensures predictable optimization behavior which NSE with static reference tour fails to deliver.
    
A key outcome concerns the comparison with exact optimization. While GLPK successfully solved small and medium instances, it failed to construct any feasible tour for 4 out of 13 instances within the 4-hour limit. This is not a matter of solution quality—it is a structural scalability barrier: the solver could not build a valid routing structure for larger problem sizes. FL-NSE, by contrast, consistently identified feasible solutions across all test cases, outperforming GLPK on 8 of the 9 comparable instances. These findings suggest that for large-scale preference-aware routing, the question is no longer how closely a metaheuristic approximates exact optimality, but whether exact methods can deliver any feasible operational plan within practical time constraint. The adaptive fuzzy mechanism appears essential to maintaining this feasibility where standard branch-and-bound exploration stalls.
	
More surprising was the behavior of the two objectives together. We entered this study expecting the standard trade-off: improving service quality would come at the expense of additional travel distance. Instead, both objectives improved on all $13$ instances. On $12$ instances, distance and dissatisfaction decreased simultaneously; on \texttt{eil76}, dissatisfaction decreased substantially ($619.93$ versus $669.67$) while distance remained effectively unchanged ($1285.52$ versus $1286.16$). The implication is that classical NSE was not trading off optimally; it was settling into local optima that were globally poor. 
	
The runtime profile introduces a compelling dynamic. Small problem instances incur a noticeable fuzzy overhead (up to $2.2\times$). Unexpectedly, larger instances run faster, dropping to $0.9\times$ beyond $150$ cities. If confirmed across a broader sample of large-scale instances, we interpret this runtime reduction as the hidden cost of stagnation becoming visible: a static reference tour wastes generations on local tweaks that yield no improvement, whereas periodic reorientation eliminates that unnecessary computation. For practical logistics applications routinely exceeding the scales tested here, this favourable scaling trend is not merely advantageous, it is decisive.
	
The comparison with distance-optimal tours raises a methodological concern that extends beyond this study. The optimal tour for \texttt{eil51} incurs a dissatisfaction of $428$, a penalty that FL\-NSE reduces by $53.5\%$ (to $199$), the largest dissatisfaction reduction observed across all instances. For \texttt{pr144}, the distance gap reaches $640\%$ (from $58,537$ to $433,026$), while still leaving a dissatisfaction of $2,120$, far below the optimal tour's penalty of $3,053$. These gaps do not indicate algorithmic deficiency; they reveal the inadequacy of distance-optimal tours as benchmarks for preference-aware routing. The TSPTW literature has established that hard temporal constraints force radical departures from shortest routes. Our results indicate that soft preference constraints generate a comparable effect: as problem scale increases, the distance-optimal tour becomes progressively decoupled from operational utility. On \texttt{pr144}, a tour that is mathematically shortest leaves customers so dissatisfied that its practical value is questionable. This decoupling is not a tension to be managed but a structural divergence that renders pure distance minimization unsuitable as a sole objective in service-oriented applications.
	
Finally, we must acknowledge what this study does not establish. First, a key structural limitation lies in the weighted sum aggregation used in our work. By scalarizing multiple criteria into a single objective, this formulation inherently fails to identify trade-off solutions located in non-convex regions of the Pareto front. Consequently, certain high-quality candidate solutions with desirable multi-objective balances may be unfairly evaluated as inferior, thereby restricting the search mechanism from fully capturing the underlying trade-off surface. Second, relying on random preferences keeps the experiment clean but misses realistic structures: spatially or temporally correlated patterns, such as clustered demand zones or time-dependent service priorities, may induce preference distributions that differ substantially from the uniform assumption. Third, while testing up to 200 cities is sufficient as a proof of concept, it remains modest for some contemporary situations.
    
These gaps set the boundaries of what can be claimed. Within them, the evidence is clear: dynamic reference tour management addresses a genuine structural limitation in NSE. It preserves feasibility, adds interpretable decision logic, and demonstrably expands the reachable solution space. For researchers working on preference-aware routing, and for practitioners who must choose between efficiency and service quality, this offers a basis that is theoretically coherent and empirically supported, a basis we intend to extend.
	
\section{Conclusion}
\label{sec:conclusion}
	
This paper makes two contributions. First, it formulates the { service oriented TSP} as a bi-objective optimization problem, where each customer is associated to a visit rank and deviations incur dissatisfaction costs, a formulation not previously addressed in the literature to the best of our knowledge. Second, it proposes an adaptive FL-NSE framework that dynamically updates the reference tour  upon stagnation detection, resolving the structural limitation of classical NSE's dependence on a single static reference tour.
	
Experimental evaluation on $13$ TSPLIB instances demonstrates consistent fitness improvements over classical NSE, with simultaneous reductions in both travel distance and customer dissatisfaction. The comparison with distance-optimal tours reveals dissatisfaction reductions reaching $53.5\%$, confirming that pure distance minimization is inadequate for service-oriented routing. On large instances, the adaptive mechanism achieves faster execution than classical NSE ($0.9\times$  beyond $150$ cities), suggesting favourable scalability.
	
Future work will extend the framework to a pure Pareto optimization scheme, bypassing the structural boundaries of scalarized scoring to effectively explore non-convex trade-off surfaces. Future efforts will also evaluate correlated preference distributions that reflect real-world logistics constraints such as spatially clustered demand.

\printcredits
\section*{Data availability}
Data available on demand. The benchmark instances are publicly available from TSP-LIB.

\section*{Declaration of competing interest}
The authors declare that they have no known competing financial interests or personal relationships that could have appeared to influence the work reported in this paper.

\section*{Funding}
This research received no specific grant from any funding agency in the 
public, commercial, or not-for-profit sectors.

\bibliographystyle{apalike}

\bibliography{refs}



\end{document}